\documentclass[showpacs,floatfix,superscriptaddress]{revtex4}
\usepackage{graphicx}
\usepackage{epsfig}
\usepackage{bm}
\usepackage[utf8]{inputenc}
\usepackage{amssymb}
\usepackage{float}
\usepackage{amsmath}
\usepackage{dcolumn}
\usepackage{float}
\usepackage{latexsym}
\usepackage{subfig}
\usepackage{tcolorbox}
\usepackage{amsthm}
\usepackage{booktabs}
\usepackage{rotating}
\usepackage{framed}
\usepackage{caption}
\usepackage[font=footnotesize,labelfont=bf]{caption}

\usepackage{cancel}
\usepackage{mathtools}

\usepackage[colorlinks]{hyperref}
\usepackage[dvipsnames]{xcolor}
\tcbuselibrary{breakable}
\newtcolorbox{revisionhighlight}{
  breakable,
  colback=yellow!35,
  colframe=yellow!35,
  boxrule=0pt,
  arc=0pt,
  left=1pt,
  right=1pt,
  top=1pt,
  bottom=1pt,
  boxsep=0pt,
  before skip=2pt,
  after skip=2pt
}

\hypersetup{
     breaklinks=true,
    pdfstartview={FitH},    % fits the width of the page to the window
    colorlinks=true,       % false: boxed links; true: colored links
    linkcolor=blue,          % color of internal links
    citecolor=red,        % color of links to bibliography
    filecolor=magenta,      % color of file links
    urlcolor=blue,           % color of external links
    anchorcolor=green,      % Color for anchor text
    linktocpage=true
}

\def\doi{http://doi.org}

\begin{document}

\title{Effective reheating in Gauss--Bonnet inflation with $\mu(\phi,X)$ coupling}

\author{Ali Seidabadi }
\email[]{aseidabadii@gmail.com}
\affiliation{Department of Theoretical Physics, Faculty of Sciences, University of Mazandaran,\\
47416-95447, Babolsar, Iran}

\author{Sara Saghafi}
\email[]{saghafisara1366@gmail.com}
\affiliation{Department of Theoretical Physics, Faculty of Sciences, University of Mazandaran,\\
47416-95447, Babolsar, Iran}

\author{Kourosh Nozari}
\email[]{nozari7450@gmail.com, knozari@umz.ac.ir  (Corresponding Author) }
\affiliation{Department of Theoretical Physics, Faculty of Sciences, University of Mazandaran,\\
47416-95447, Babolsar, Iran}

\begin{abstract}
We study effective reheating in a scalar--Gauss--Bonnet
inflationary model with a phase-space-dependent coupling
$\mu(\phi,X)$, in which a compact field-space feature is combined
with a bounded kinetic gate. The modified inflationary background
determines the pivot-scale quantities and the effective energy
density at the end of inflation. These quantities are then used to
derive the reheating duration $N_{\rm re}$ and temperature
$T_{\rm re}$ through the thermal-history matching relation.
We first perform two fixed-pivot reference scans by varying the
overall Gauss--Bonnet strength $\lambda_{\rm GB}$ and the kinetic-gate amplitude
$g_{_X}$ separately. For the reference parameter choices,
increasing either parameter increases $N_{\rm re}$ and decreases
$T_{\rm re}$ for the selected reheating equations of state.
Additional benchmark calculations clarify how these variations
depend on the dynamical regime of the model. In the
$\lambda_{\rm GB}$ scan, the increase in $N_{\rm re}$ and the
decrease in $T_{\rm re}$ persist, although both variations become
strongly suppressed when the coupling is more localized or when the
end of inflation is controlled more strongly by the E-model
potential. In the $g_{_X}$ scan, stronger field-space localization and
kinetic saturation can instead lead to a slight decrease in
$N_{\rm re}$ and an increase in $T_{\rm re}$ as $g_{_X}$ is increased.
When the bounded kinetic contribution is considered together with
a weaker overall Gauss--Bonnet interaction, the resulting changes in the
reheating quantities become nearly negligible.
The fixed-pivot predictions of the representative and alternative
benchmarks are compared with CMB constraints.These reheating constraints are then discussed for four representative
values of the effective equation-of-state parameter,
$\overline{w}_{\rm re}=-1/3,0,2/3,$ and $1$. Finally, the analysis is
generalized to a time-dependent reheating equation of state
$w_{\rm re}(N)$, showing that the thermal-history matching relations
depend only on its e-fold average $\overline{w}_{\rm re}$, while the
detailed evolution of $w_{\rm re}(N)$ remains degenerate within the
effective framework adopted here.
\end{abstract}

\maketitle
\noindent\textbf{Keywords:}
Inflation, Gauss--Bonnet gravity, phase-space coupling.

\enlargethispage{\baselineskip}
\tableofcontents

\section{Introduction}
\label{sec:introduction}

Inflation gives a simple explanation for the large-scale uniformity of the
Universe and for the origin of primordial perturbations. It does not, however,
complete the transition to the hot radiation era. The energy stored in the
inflaton must be transferred to other degrees of freedom and the produced
particles must approach a thermal state. This post-inflationary stage is known
as reheating. Early work already showed that reheating is needed to connect an
inflationary phase to the later thermal history of the Universe
\cite{AlbrechtSteinhardtTurnerWilczek1982}. It was later found that the transfer
of energy can include a rapid nonperturbative stage driven by parametric
resonance, usually called preheating
\cite{KofmanLindeStarobinsky1994,KofmanLindeStarobinsky1997}.

Nonperturbative amplification of inflaton fluctuations during preheating can
also lead to fragmentation of the homogeneous condensate and to the formation
of localized, long-lived configurations such as oscillons. Recent
numerical-relativity studies have shown that such structures may also arise in
the presence of a scalar--Gauss--Bonnet interaction. For sufficiently weak
higher-curvature corrections the oscillon properties can remain close to
those obtained in general relativity, whereas for stronger couplings the large
local curvatures generated during oscillon formation can invalidate the
effective-field-theory description and make homogeneous treatments
unreliable~\cite{WaemingEtAl2026}.
The main physical mechanisms and their particle-physics implications have been
reviewed in Refs.~\cite{AllahverdiEtAl2010,AminEtAl2015}.

The detailed microphysics of reheating is model dependent. A useful first
step is therefore to describe the expansion during this era by an average
equation-of-state parameter \(w_{\rm re}\). Using this parametrization, the number of
reheating e-folds \(N_{\rm re}\) and the reheating temperature \(T_{\rm re}\) can
be related to the time at which a CMB pivot mode left the horizon. This relation
was developed and applied to several inflationary models in
Refs.~\cite{DaiKamionkowskiWang2014,MartinRingevalVennin2015,
MunozKamionkowski2015,CookEtAl2015}. Nonlinear studies also show that the time
needed to reach radiation domination depends on the form of the inflaton
potential near its minimum \cite{LozanovAmin2017}. Detailed analyses of neutrino thermalization and primordial
light-element abundances in low-temperature reheating scenarios require
the reheating temperature to be at least a few MeV
\cite{Kawasaki:2000en,deSalasEtAl2015,Hasegawa:2019jsa,Kawasaki:1999na}.
Accordingly, we adopt $10\,\mathrm{MeV}$ as a conservative lower bound
on $T_{\rm re}$.

The Gauss--Bonnet term is a well-known higher-curvature combination. It appears
in the low-energy expansion of string-inspired gravitational actions
\cite{Zwiebach1985,GrossSloan1987,Nojiri:2010wj,Nojiri:2017ncd}. In four dimensions a constant coefficient
of the Gauss--Bonnet invariant does not change the field equations, but a
coupling to a scalar field makes it dynamical. Scalar--Gauss--Bonnet cosmology
was studied, for example, in Ref.~\cite{NojiriOdintsovSasaki2005}. During
inflation, this coupling can change both the background evolution and the
primordial perturbations.

Several early studies derived the scalar and tensor spectra in
Gauss--Bonnet inflation and examined the stability of the perturbations
\cite{SatohKannoSoda2008,SatohSoda2008,GuoSchwarz2009,Kanti:2015pda}. The slow-roll hierarchy
and the modified relation between the tensor amplitude and tensor tilt were
worked out in Refs.~\cite{GuoSchwarz2010,Satoh2010}. The Gauss--Bonnet coupling
can therefore change \(n_s\), \(r\), and the tensor tilt without changing the
inflaton potential alone. More general second-order scalar--tensor theories,
including the Gauss--Bonnet interaction as a special sector, were developed in
Ref.~\cite{KobayashiYamaguchiYokoyama2011} and reviewed in
Ref.~\cite{Kobayashi2019Review}.

The potential used in the present work belongs to the E-model
\(\alpha\)-attractor family \cite{KalloshLindeRoest2013}. The effect of a
field-dependent Gauss--Bonnet coupling on familiar inflationary potentials was
compared with CMB data in Refs.~\cite{JiangHuGuo2013,KohEtAl2014}. Analytic
relations between the potential, the coupling function, and the inflationary
observables have also been studied in slow-roll and constant-roll examples
\cite{YiGongSabir2018}. These results show that the form and strength of the
coupling must be chosen together with the potential; changing the coupling can
move a model in the \((n_s,r)\) plane, but it can also introduce instabilities or
alter the end of inflation.

This last point is important for reheating. In some Gauss--Bonnet models the
late stage coupling prevents the usual oscillatory phase of the inflaton. This
problem was shown for an inverse-power coupling, and a modified coupling that
allows perturbative reheating was introduced in
Ref.~\cite{vanDeBruckDimopoulosLongden2016}. Reheating and unitarity were then
used to further restrict scalar--Gauss--Bonnet inflation
\cite{BhattacharjeeMaityMukherjee2017}. The reheating parameters were also
related to the primordial gravitational-wave spectrum
\cite{KohLeeTumurtushaa2018}. Reheating was studied for canonical and
noncanonical scalar fields with several Gauss--Bonnet couplings and potentials,
including E- and T-models \cite{RashidiNozari2020}. A model containing both
nonminimal kinetic and Gauss--Bonnet couplings was also studied, including its
perturbation equations and reheating behavior
\cite{GrandaJimenez2021}. Gravitational-wave production and preheating
parameters were studied for a power-law Gauss--Bonnet model in
Ref.~\cite{ElBourakadiEtAl2021}.

More recent papers have followed the evolution closer to the end of inflation.
In parallel, GW170817-compatible Einstein--Gauss--Bonnet inflationary
frameworks have been developed and progressively refined by imposing the
observational constraint on the tensor propagation speed, with recent
extensions also examining the reheating-era evolution of the
Gauss--Bonnet coupling and the viability of these models in light of
ACT data
\cite{Odintsov:2020sqy,Oikonomou:2021kql,Oikonomou:2024etl,
Odintsov:2025wai,Yogesh:2026esn}. An inflation-to-reheating evolution with scalar decay was constructed, and the
stability of the reheating solution was tested in
Ref.~\cite{OdintsovPaul2023}. The change in the Gauss--Bonnet coupling between
inflation, the end of inflation, and reheating was studied in a
GW170817-compatible theory \cite{OikonomouTsybaRazina2024}. After the release
of ACT DR6, the effect of the updated CMB range on the reheating bounds in
scalar--Einstein--Gauss--Bonnet gravity was examined in
Ref.~\cite{OdintsovPaul2025}.

The observational comparison in this paper uses the inflationary constraints
from Planck 2018, BICEP/Keck 2018, and ACT DR6
\cite{Planck2018Inflation,BKKeck2021,ACTDR6Louis2025}. These data now probe a
small region of the \((n_s,r)\) plane and make the assumed reheating history more
important. Recent examples include a Gauss--Bonnet model with a Higgs potential
tested against ACT DR6 \cite{AhghariFarhoudi2026}, and an analysis of E- and
T-model \(\alpha\)-attractors in which ACT DR6 was used to constrain both
inflation and reheating \cite{HaquePalPaul2026}.

In our previous paper, we replaced the usual field coupling
\(f(\phi)\mathcal G\) by a phase-space coupling \(\mu(\phi,X)\mathcal G\)
\cite{SeidabadiSaghafiNozari2026}. The coupling was localized in field space by
a compact bump and modulated by a bounded kinetic gate. We solved the modified
background equations, checked ghost and gradient stability in the scalar and
tensor sectors, and studied how the overall Gauss--Bonnet strength and the
kinetic gate change the pivot-scale observables. That work was focused on the
inflationary era and did not discuss the reheating history.
The motivation for introducing the phase-space-dependent coupling $\mu(\phi,X)$, rather than restricting the Gauss--Bonnet interaction
to the commonly considered form $f(\phi)\mathcal{G}$, is to allow the
higher-curvature sector to respond not only to the position of the
inflaton in field space but also to its instantaneous kinetic regime.
The homogeneous scalar background evolves along a trajectory in the
two-dimensional phase space $\{\phi,X\}$, with $X=-\frac{1}{2}g^{\mu\nu}\partial_{\mu}\phi\,\partial_{\nu}\phi $.

A purely field-dependent coupling probes only the $\phi$ coordinate
of this trajectory, whereas a coupling $\mu(\phi,X)$ can distinguish
dynamical regimes having different kinetic properties even when their
field-space dependence is similar. In this sense, the $X$ dependence
provides an additional dynamical modulation of the Gauss--Bonnet
interaction rather than merely a redefinition of its overall strength.

In the parametrization adopted here, the two dependences have distinct
roles. The compact bump localizes the modification in field space,
whereas the function $G(X)$ acts as a bounded kinetic gate. The latter
allows the Gauss--Bonnet correction to become sensitive to the kinetic
state of the inflaton while approaching a finite saturation value at
large $X$. This construction therefore provides a controlled
phenomenological extension of the standard scalar--Gauss--Bonnet
coupling. Importantly, the usual field-dependent case is contained as
a limiting sector of the model when the explicit kinetic modulation is
switched off. Our purpose is consequently not to regard $\mu(\phi,X)$
as a replacement for $f(\phi)$ in general, but to examine whether an
additional, bounded dependence on the scalar kinetic sector can lead
to distinguishable inflationary and reheating phenomenology.

The present paper continues that analysis. We keep the full coupling in the
form \(\lambda_{\rm GB}\mu(\phi,X)\) and use the modified background equations to
find the effective energy density at the end of inflation. We then obtain
\(N_{\rm re}\) and \(T_{\rm re}\) from the thermal-history matching relation. Two fixed-pivot reference scans are used to examine $\lambda_{\rm GB}$ and $g_{_X}$ separately. Additional benchmark tests
are performed to determine whether the corresponding reheating
trends persist when the potential shape, the field-space
localization, the kinetic saturation scale, and the overall
Gauss--Bonnet strength are changed. The resulting predictions for
$n_s$ and $r$ are compared with the CMB contours. We then study the reheating constraints of two
representative and two alternative parameter sets for
$w_{\rm re}=-1/3$, $0$, $2/3$, and $1$, considering the adopted BBN
temperature bound on each branch.

In general, the reheating equation-of-state parameter is not expected
to remain constant throughout the entire post-inflationary evolution.
The coherent oscillation of the inflaton, particle production, and the
gradual emergence of a radiation component can lead to a
time-dependent equation of state $w_{\rm re}(N)$
\cite{SahaAnandSriramkumar2020,GarciaPierre2023,GhoshGhosh2025}.
For the present thermal-history matching analysis, the representative
values employed numerically are interpreted as values of the
e-fold-averaged parameter
$\overline{w}_{\rm re}
=N_{\rm re}^{-1}\int_{0}^{N_{\rm re}}w_{\rm re}(N)\,dN$.
The detailed generalization is presented in
Section.~\ref{subsec:varying-eos}.

The paper is organized as follows. Section~\ref{sec:model} introduces
the model, the modified background equations, the inflationary
observables, and the effective reheating relations. The fixed-pivot
reference scans and the corresponding alternative benchmark tests
are presented in
Subsec.~\ref{subsec:fixed-pivot-parameter-scans}. The CMB predictions
of the representative and alternative benchmarks, together with
their reheating constraints, are discussed in
Subsec.~\ref{subsec:cmb-and-reheating}. The generalization to a
time-dependent reheating equation of state and the interpretation of
the numerical results in terms of the e-fold-averaged parameter
$\overline{w}_{\rm re}$ are presented in
Section.~\ref{subsec:varying-eos}. Finally,
Sec.~\ref{sec:conclusions} summarizes the main results.

%=====================================================
\section{The model and effective reheating framework}
\label{sec:model}
We consider a scalar--Gauss--Bonnet inflationary model in which the
coupling depends on both the scalar field and its kinetic term. The
action is written as
\begin{equation}
S=\int d^4x \sqrt{-g}\left[
\frac{1}{2}R - \frac{1}{2}g^{\mu\nu}\partial_\mu\phi\,\partial_\nu\phi - V(\phi)
-\lambda_{GB}\mu(\phi,X)\,\mathcal{G}
\right].
\label{eq:action}
\end{equation}
Here \(\mathcal G\) is the Gauss--Bonnet invariant,
\begin{equation}
\mathcal{G} = R^{\mu\nu\rho\sigma}R_{\mu\nu\rho\sigma}
- 4R^{\mu\nu}R_{\mu\nu}
+ R^2 ,
\label{eq:GBinv}
\end{equation}
and the kinetic term is denoted by
\begin{equation}
X \equiv -\frac{1}{2}g^{\mu\nu}\partial_\mu\phi\,\partial_\nu\phi.
\label{eq:Xdef}
\end{equation}
Also, \(R\) is the Ricci scalar, \(\phi\) denotes the inflaton, and
\(\lambda_{\rm GB}\) sets the overall strength of the
\(\mu(\phi,X)\mathcal G\) interaction. Throughout this work, the full coupling is
\(\lambda_{\rm GB}\mu(\phi,X)\) and we use reduced Planck units,
\(M_{\rm Pl}=1\).
For a spatially flat FLRW metric with a general lapse function 
\begin{align}
ds^2=-N^2(t)dt^2+a^2(t) d\vec{x}^{\,2},
\label{eq:7}
\end{align}
the background equations are 

\begin{equation}
3H^2
=
\frac{1}{2}\dot{\phi}^{\,2}
+V(\phi)
-24\lambda_{\rm GB}H^3
\left(
\mu_{,\phi}\dot{\phi}
+\mu_{,X}\dot{\phi}\ddot{\phi}
\right)
+48\lambda_{\rm GB}H^4X\mu_{,X}.
\label{eq:Friedmann}
\end{equation}

\begin{align}
-2\dot H
={}&
\dot{\phi}^{\,2}
+8\lambda_{\rm GB}H^2
\left(
\mu_{,\phi\phi}\dot{\phi}^{\,2}
+\mu_{,\phi}\ddot{\phi}
\right)
+16\lambda_{\rm GB}H\dot H\,
\mu_{,\phi}\dot{\phi}
\nonumber\\
&-8\lambda_{\rm GB}H^3\mu_{,\phi}\dot{\phi}
+16\lambda_{\rm GB}H^2
\left(2\dot H-H^2\right)Y
+16\lambda_{\rm GB}H^3\dot Y .
\label{eq:Raychaudhuri}
\end{align}

\begin{equation}
\left(
1-\lambda_{\rm GB}\mu_{,X}\mathcal G
\right)
\left(
\ddot{\phi}+3H\dot{\phi}
\right)
-\lambda_{\rm GB}
\frac{d}{dt}
\left(
\mu_{,X}\mathcal G
\right)\dot{\phi}
+V_{,\phi}
+\lambda_{\rm GB}\mu_{,\phi}\mathcal G
=0.
\label{eq:scalar-field}
\end{equation}

where
\(\mu_{,\phi}\equiv\partial\mu/\partial\phi\),
\(\mu_{,\phi\phi}\equiv\partial^2\mu/\partial\phi^2\),
\(\mu_{,X}\equiv\partial\mu/\partial X\), and
\(Y\equiv X\mu_{,X}\). Moreover,
\(V_{,\phi}\equiv\partial V/\partial\phi\).

The distinction from a purely field-dependent coupling is explicit in
Eqs.~\eqref{eq:Friedmann}--\eqref{eq:scalar-field}. For $\mu_{,X}=0$,
all terms generated by the kinetic dependence disappear and the usual
field-dependent scalar--Gauss--Bonnet sector is recovered. For
$\mu_{,X}\neq0$, the kinetic gate modifies both the effective background
energy density and the scalar-field equation of motion. Thus, the parameter
$g_{_X}$ does not simply rescale the Gauss--Bonnet interaction, but controls a
genuinely distinct dynamical contribution.

Since \(\lambda_{\rm GB}\) is constant, the derivative term appearing in
Eq.~\eqref{eq:scalar-field} can be written as
\begin{equation}
\lambda_{\rm GB}
\frac{d}{dt}\left(\mu_{,X}\mathcal G\right)
=
\lambda_{\rm GB}
\left(
\dot{\mu}_{,X}\mathcal G
+
\mu_{,X}\dot{\mathcal G}
\right),
\end{equation}
where
\begin{equation}
\dot{\mathcal G}
=
24\left[
2H\dot H\left(\dot H+H^2\right)
+
H^2\left(\ddot H+2H\dot H\right)
\right].
\end{equation}
\\
The first two slow-roll parameters are defined by
\begin{equation}
\epsilon_1 \equiv -\frac{\dot{H}}{H^2}, \qquad
\epsilon_2 \equiv \frac{\dot{\epsilon}_1}{H\epsilon_1}.
\label{eps1}
\end{equation}
In the presence of the Gauss--Bonnet contribution, one can introduce the Gauss--Bonnet slow-roll hierarchies
\begin{equation}
\delta_1 \equiv -8\lambda_{\rm GB}H\dot{\mu}, \qquad
\delta_2 \equiv \frac{\dot{\delta}_1}{H\delta_1}, \qquad
\kappa_1 \equiv 16\lambda_{\rm GB}H^2X\mu_{,X}, \qquad
\kappa_2 \equiv \frac{\dot{\kappa}_1}{H\kappa_1}.
\label{eq:deltakappa-def}
\end{equation}
Throughout this work, we adopt the potential as the E-model 
\begin{equation}
V(\phi)=V_0\left(1-e^{-b\phi}\right)^2,
\qquad
b=\sqrt{\frac{2}{3\alpha}} \, .
\label{eq:potential}
\end{equation}
Here \(V_0\) fixes the inflationary energy scale, while \(\alpha\)
controls the curvature of the potential.

Before specifying the coupling, we emphasize that the dependence on
$X$ is introduced as a controlled extension of the conventional
field-dependent Gauss--Bonnet interaction. In the limit in which the
kinetic modulation is switched off, the model reduces to a coupling
depending only on $\phi$. The purpose of the $X$ sector is therefore
to introduce an additional dependence on the dynamical state of the
inflaton while keeping this correction bounded.
For clarity, we distinguish between the constant overall
Gauss--Bonnet strength $\lambda_{\rm GB}$ and the phase-space-dependent
coupling function $\mu(\phi,X)$. The complete coefficient multiplying
the Gauss--Bonnet invariant in the action is
$
F_{\rm GB}(\phi,X)\equiv
\lambda_{\rm GB}\mu(\phi,X).
$
Here, $\lambda_{\rm GB}$ controls the overall amplitude of the
Gauss--Bonnet interaction, whereas $\mu(\phi,X)$ determines its
dependence on the scalar-field position and kinetic state. Thus,
$\mu(\phi,X)$ is not an additional coupling independent of
$\lambda_{\rm GB}\mu(\phi,X)$; it is the phase-space-dependent function
entering the full Gauss--Bonnet coefficient.

Up to this point, $\mu(\phi,X)$ has been kept as a general
phase-space-dependent function in order to write the background
equations in a form independent of a particular parametrization. For
the numerical analysis, we now specify the functional form of the same
$\mu(\phi,X)$ appearing in Eq.~\eqref{eq:action}. Following
Ref.~\cite{SeidabadiSaghafiNozari2026}, we adopt

\begin{equation}
\mu(\phi,X)=e^{-\alpha_\mu \phi}\Bigl[1+A\,B\!\Bigl(\frac{\phi-\phi_\star}{\Delta}\Bigr)\Bigr]
\Bigl[1+g_{{_X}}\,G(X)\Bigr].
\label{eq:muModel}
\end{equation}

Accordingly, the full coefficient of $\mathcal{G}$ used throughout
the numerical analysis is $\lambda_{\rm GB}\mu(\phi,X)$, with
$\lambda_{\rm GB}$ controlling its overall strength. Within
$\mu(\phi,X)$, the parameter $g_X$ controls the amplitude of the
explicit bounded kinetic modulation $G(X)$; setting $g_X=0$ removes
this explicit kinetic dependence while retaining the field-dependent
part of the coupling.

The bump function \(B\) confines the field-dependent feature to a
finite region around \(\phi_\star\), while \(G(X)\) provides a bounded
kinetic modulation. The parameters \(A\), \(\alpha_\mu\), and $g_{_X}$
control the bump amplitude, the smooth field dependence, and the kinetic
contribution, respectively.
Accordingly, the bump $B$ as a $C^\infty$ compact-support function is written as 
\begin{equation}
B(z)=
\begin{cases}
\exp\!\left[-\dfrac{1}{1-z^{2}}\right], & |z|<1,\\[6pt]
0, & |z|\ge 1,
\end{cases}
\qquad
z\equiv\frac{\phi-\phi_\star}{\Delta}.
\label{eq:bump}
\end{equation}
Also, the kinetic sector is defined by
\begin{equation}
G(X)\equiv \frac{u(X)}{1+\beta_{\rm GX}\,u(X)}.
\label{eq:gate}
\end{equation}
The dimensionless variable \(u(X)\) is defined as
\[
u(X)\equiv \left(\frac{X}{M^{4}}\right)^p ,
\]
where the scale \(M\) determines where the kinetic gate becomes active,
\(p>0\) controls the steepness of the transition, and
\(\beta_{GX}>0\) fixes the limiting value
\(G(X)\rightarrow1/\beta_{GX}\).

The bounded form of $G(X)$ is particularly useful in this respect:
for small $X/M^4$ the kinetic correction can remain suppressed,
while for sufficiently large $X/M^4$ it approaches a finite value,
$
G(X)\longrightarrow \frac{1}{\beta_{GX}},$
thereby avoiding an unbounded enhancement of the Gauss--Bonnet
interaction from the kinetic sector.

The inflationary observables are evaluated using the scalar and tensor
power spectra in the slow-roll approximation,
\begin{equation}
\mathcal{P}_{\mathcal R}(k)
=
\left.
\frac{H^2}{8\pi^2Q_sc_s^3}
\right|_{c_sk=aH},
\qquad
\mathcal{P}_{t}(k)
=
\left.
\frac{H^2}{2\pi^2Q_tc_t^3}
\right|_{c_tk=aH}.
\label{eq:power-spectra}
\end{equation}
Here \(Q_s\) and \(Q_t\) are the scalar and tensor coefficients,
while \(c_s\) and \(c_t\) are the corresponding propagation speeds,
evaluated on the full \(\mu(\phi,X)\) background. The absence of ghost and gradient instabilities requires
\begin{equation}
    Q_s>0,\qquad Q_t>0,\qquad c_s^2>0,\qquad c_t^2>0.
\label{eq:stability}
\end{equation}
The explicit expressions for these quantities in the
phase-space-dependent Gauss--Bonnet model, together with the
corresponding stability analysis, were derived in our previous
work~\cite{SeidabadiSaghafiNozari2026}.
In the present analysis, only numerical solutions satisfying the
same stability conditions at the pivot scale are retained.
The pivot time is defined by
\begin{equation}
N_\star=N_{\rm end}-N_{\rm pivot}.
\label{eq:pivot-time}
\end{equation}
The comoving wavenumber associated with the background evolution is
\begin{equation}
k(N)=\frac{a(N)H(N)}{c_s(N)},
\qquad
\frac{d\ln k}{dN}
=
1+\frac{d\ln H}{dN}
-\frac{d\ln c_s}{dN}.
\label{eq:k-of-N}
\end{equation}
The scalar spectral index and the tensor-to-scalar ratio are therefore
evaluated as
\begin{equation}
n_s-1
=
\left.
\frac{d\ln\mathcal{P}_{\mathcal R}/dN}
     {d\ln k/dN}
\right|_{N=N_\star},
\qquad
r
=
\left.
\frac{\mathcal{P}_{t}}
     {\mathcal{P}_{\mathcal R}}
\right|_{N=N_\star}.
\label{eq:ns-r}
\end{equation}

The inflationary background is obtained by solving the modified
equations for the chosen potential and coupling. The resulting
trajectory is described by \(\{\phi(N),X(N)\}\), while the
Gauss--Bonnet contribution is controlled by the field-space bump and
the kinetic gate. Inflation ends at \(N=N_{\rm end}\), where
\(\epsilon_1=1\). The background at this time determines the effective
energy density \(\rho_{\rm end}\), which provides the initial condition
for the reheating stage.
Accordingly, the modified Hamiltonian constraint can be written as
\begin{equation}
3H^2=\rho_{\rm eff},
\label{eq:hamiltonian-effective-density}
\end{equation}
where
\begin{equation}
\rho_{\rm eff}
=
X+V(\phi)
-24\lambda_{\rm GB}H^3
\left(
\mu_{,\phi}\dot\phi+\mu_{,X}\dot X
\right)
+48\lambda_{\rm GB}H^4X\mu_{,X}.
\label{eq:rho-eff}
\end{equation}
The initial energy density of the reheating stage is therefore defined as
\begin{equation}
\rho_{\rm end}\equiv \rho_{\rm eff}(N_{\rm end}),
\label{eq:rho-end-def}
\end{equation}
or explicitly
\begin{equation}
\rho_{\rm end}
=
\left[
X+V(\phi)
-24\lambda_{\rm GB}H^3
\left(
\mu_{,\phi}\dot\phi+\mu_{,X}\dot X
\right)
+48\lambda_{\rm GB}H^4X\mu_{,X}
\right]_{N=N_{\rm end}} .
\label{eq:rho-end-explicit}
\end{equation}
To describe the reheating stage, we treat the post-inflationary universe as dominated by an effective fluid
with a constant equation of state parameter, 
\begin{equation}
p_{\rm re}=w_{\rm re}\rho_{\rm re},
\qquad
w_{\rm re}=\mathrm{constant}.
\label{eq:constant-wre}
\end{equation}
Accordingly, the effective pressure follows the modified Raychaudhuri equation,
\begin{equation}
\rho_{\rm eff}+p_{\rm eff}
=
-2\dot H .
\label{eq:rho-plus-p-eff}
\end{equation}
Therefore,
\begin{equation}
p_{\rm eff}
=
-2\dot H-\rho_{\rm eff},
\label{eq:p-eff-raychaudhuri}
\end{equation}
\begin{align}
p_{\rm eff}
=&\;
X-V(\phi)
+8\lambda_{\rm GB}H^{2}
\left(
\mu_{,\phi\phi}\dot{\phi}^{2}
+
\mu_{,\phi}\ddot{\phi}
\right)
+16\lambda_{\rm GB}H\dot H\,
\mu_{,\phi}\dot{\phi}
\nonumber\\
&\;
+16\lambda_{\rm GB}H^{3}\mu_{,\phi}\dot{\phi}
+24\lambda_{\rm GB}H^{3}\mu_{,X}\dot X
\nonumber\\
&\;
+16\lambda_{\rm GB}H^{2}
\left(2\dot H-H^{2}\right)Y
+16\lambda_{\rm GB}H^{3}\dot Y
-48\lambda_{\rm GB}H^{4}Y .
\label{eq:p-eff-muXGB}
\end{align}

The total effective fluid obeys
\begin{equation}
\dot\rho_{\rm eff}
+
3H
\left(
\rho_{\rm eff}+p_{\rm eff}
\right)
=
0 .
\label{eq:total-effective-continuity}
\end{equation}
For a constant reheating equation of state, the continuity equation
becomes
\begin{equation}
\dot\rho_{\rm re}
+
3H(1+w_{\rm re})\rho_{\rm re}
=
0 .
\label{eq:reheating-cont-time}
\end{equation}
Using \(dN=Hdt\), the continuity equation can be written as
\begin{equation}
\frac{d\rho_{\rm re}}{dN}
+
3(1+w_{\rm re})\rho_{\rm re}
=
0 .
\label{eq:reheating-cont-N}
\end{equation}
Since \(w_{\rm re}\) is constant here, Eq.~\eqref{eq:reheating-cont-N} integrates to
\begin{equation}
\rho_{\rm re}
=
\rho_{\rm end}
e^{-3(1+w_{\rm re})N_{\rm re}},
\label{eq:rhore}
\end{equation}
where
\(N_{\rm re}\equiv\ln(a_{\rm re}/a_{\rm end})\)
is the duration of the reheating stage.
Using Eq.~\eqref{eq:rho-end-explicit}, Eq.~\eqref{eq:rhore} can be written in the form
\begin{align}
\rho_{\rm re}
={}&
\Big[
X+V
-24\lambda_{\rm GB}H^3
\left(
\mu_{,\phi}\dot{\phi}
+\mu_{,X}\dot X
\right)
+48\lambda_{\rm GB}H^4X\mu_{,X}
\Big]_{N=N_{\rm end}}
\nonumber\\
&\times
\exp\left[-3(1+w_{\rm re})N_{\rm re}\right].
\label{eq:rho-re-explicit}
\end{align}
Assuming that the Universe is thermalized at the end of reheating, one can write $\rho_{re}$ as a function of $T_{re}$,
\begin{equation}
\rho_{\rm re}
=
\frac{\pi^2}{30}g_{\rm re}T_{\rm re}^4,
\label{eq:rhoT}
\end{equation}
where \(g_{\rm re}\) is the number of relativistic degrees of freedom at reheating. Using Eqs.~\eqref{eq:rhore} and \eqref{eq:rhoT}, the reheating temperature is
\begin{equation}
T_{\rm re}
=
\left(
\frac{30}{\pi^2g_{\rm re}}
\rho_{\rm end}
\right)^{1/4}
\exp\left[
-\frac34(1+w_{\rm re})N_{\rm re}
\right].
\label{eq:Tre-Nre}
\end{equation}
Substituting the explicit expression for \(\rho_{\rm end}\) gives
\begin{align}
T_{\rm re}
={}&
\left(
\frac{30}{\pi^2g_{\rm re}}
\right)^{1/4}
\Big[
X+V
-24\lambda_{\rm GB}H^3
\left(
\mu_{,\phi}\dot{\phi}
+\mu_{,X}\dot X
\right)
+48\lambda_{\rm GB}H^4X\mu_{,X}
\Big]_{N=N_{\rm end}}^{1/4}
\nonumber\\
&\times
\exp\left[
-\frac{3}{4}(1+w_{\rm re})N_{\rm re}
\right].
\label{eq:Tre-explicit}
\end{align}
The pivot mode satisfies the sound-horizon crossing condition
\begin{equation}
c_{s\star}k_\star
=
a_\star H_\star ,
\label{eq:sound-horizon}
\end{equation}
and the ratio between the pivot scale and the present scale is obtained by following the expansion history from horizon exit to today,
\begin{equation}
k_\star
=
\frac{H_\star}{c_{s\star}}
e^{-N_{\rm pivot}}e^{-N_{\rm re}}
\frac{T_0}{T_{\rm re}}
\left(
\frac{g_{s0}}{g_{s,{\rm re}}}
\right)^{1/3} .
\label{eq:matching}
\end{equation}
Taking the logarithm of Eq.~\eqref{eq:matching} gives
\begin{equation}
N_{\rm re}
=
-N_{\rm pivot}
+
\ln\left(
\frac{H_\star}{c_{s\star}k_\star}
\right)
+
\ln T_0
+
\frac13
\ln\left(
\frac{g_{s0}}{g_{s,{\rm re}}}
\right)
-
\ln T_{\rm re} .
\label{eq:Nre-before}
\end{equation}
Equivalently, one can write
\begin{equation}
N_{\rm re}
=
\frac{4}{1-3w_{\rm re}} f,
\label{eq:Nre-final}
\end{equation}
where
\begin{align}
f
=&\;
-N_{\rm pivot}
+
\ln\left(
\frac{H_\star}{c_{s\star}k_\star}
\right)
+
\ln T_0
+
\frac13
\ln\left(
\frac{g_{s0}}{g_{s,{\rm re}}}
\right)
-
\frac14
\ln\left(
\frac{30}{\pi^2g_{\rm re}}
\right)
\nonumber\\
&-
\frac14
\ln\left\{
\left[
X+V
-24\lambda_{\rm GB}H^3
\left(
\mu_\phi\dot\phi+\mu_X\dot X
\right)
+48\lambda_{\rm GB}H^4X\mu_{,X}
\right]_{N=N_{\rm end}}
\right\} .
\label{eq:Bmatch-expanded-model}
\end{align}
Using the Hamiltonian constraint, \(f\) can be written more compactly as
\begin{align}
f
=&\;
-N_{\rm pivot}
+
\ln\left(
\frac{H_\star}{c_{s\star}k_\star}
\right)
+
\ln T_0
+
\frac13
\ln\left(
\frac{g_{s0}}{g_{s,{\rm re}}}
\right)
-
\frac14
\ln\left(
\frac{30}{\pi^2g_{\rm re}}
\right)
-
\frac14\ln\rho_{\rm end}.
\label{eq:Bmatch}
\end{align}
Once \(N_{\rm re}\) is determined, the reheating temperature follows from
\begin{equation}
T_{\rm re}
=
\left[
\frac{30}{\pi^2g_{\rm re}}
\rho_{\rm end}
e^{-3(1+w_{\rm re})N_{\rm re}}
\right]^{1/4} .
\label{eq:Tre-final}
\end{equation}

It is worth noting that, for the numerical evaluation, we adopt
$
k_\star=0.05\,{\rm Mpc}^{-1},
T_0=2.348\times10^{-13}\,{\rm GeV},
$
together with
$
g_{\rm re}=g_{s,{\rm re}}=106.75,
g_{s0}=3.91.
$
The conversion to reduced Planck units is performed using
$M_{\rm Pl}=2.435\times10^{18}\,{\rm GeV}$ and
$1\,{\rm Mpc}=1.563738\times10^{38}\,{\rm GeV}^{-1}$.

\section{Numerical results}
\label{sec:numerical-results}
The Gauss--Bonnet contribution in this model is mainly controlled by
two parameters. The parameter \(\lambda_{\rm GB}\) sets the overall Gauss--Bonnet strength, while $g_{_X}$ is the kinetic-gate amplitude that
controls the explicit kinetic modulation. Both parameters modify the inflationary background and,
through \(H_\star\), \(c_{s\star}\), and \(\rho_{\rm end}\), affect the
reheating duration and temperature.

The numerical analysis proceeds in three steps. First, two
fixed-pivot reference scans are performed to examine separately the
effects of $\lambda_{\rm GB}$ and $g_{_X}$ on the reheating parameters. Second, the scans are repeated for
alternative parameter sets in order to test the benchmark
dependence of the inferred reheating trends. Finally, the
fixed-pivot CMB predictions of the representative and alternative
benchmarks are compared, and their reheating constraints are analyzed.

\subsection{Effect of \(\lambda_{\rm GB}\) and $g_{_X}$ on reheating parameters}
\label{subsec:fixed-pivot-parameter-scans}
We examine the effects of \(\lambda_{\rm GB}\) and $g_{_X}$ through two
fixed-pivot scans. In the first scan, \(\lambda_{\rm GB}\) is varied at
fixed $g_{_X}$, while in the second scan $g_{_X}$ is varied at fixed
\(\lambda_{\rm GB}\). The parameter values used in the two scans are
listed in Table~\ref{tab:fixed-pivot-scan-setup}. Within each scan, one scan parameter is varied while the remaining parameters are held fixed.

The values of \(w_{\rm re}\) listed in
Table~\ref{tab:fixed-pivot-scan-setup} are representative choices for
which all reported scan points satisfy
\(N_{\rm re}>0\) and
\(T_{\rm re}\geq10\,{\rm MeV}\). These values are used to illustrate
how the reheating predictions depend on the assumed equation of state. Also, the potential normalization $V_0$ is calibrated separately for
each fixed-pivot solution by imposing
$A_s=2.1\times10^{-9}$ at the pivot scale.

\begin{table*}[h]
\caption{
Parameter choices used in the two fixed-pivot scans. The entries in the
\(\lambda_{\rm GB}\) and $g_{_X}$ rows indicate the values varied in the
corresponding scan; all other parameters are kept fixed.
}
\label{tab:fixed-pivot-scan-setup}

\begin{ruledtabular}
\begin{tabular}{lcc}

Parameter
&
\(\lambda_{\rm GB}\)-scan
&
$g_{_X}$-scan
\\
\hline

\(N_{\rm pivot}\)
&
\(55\)
&
\(67\)
\\

\(\lambda_{\rm GB}\)
&
\(\displaystyle
0,\;
5\times10^{-4},\;
10^{-3},\;
2\times10^{-3}
\)
&
\(0.09\)
\\

\(g_X\)
&
\(10^{-3}\)
&
\(\displaystyle
\begin{array}{c}
0,\;0.005,\\[-1mm]
0.01,\;0.02,\;0.05
\end{array}
\)
\\
\hline

\(\alpha\)
&
\(1\)
&
\(1\)
\\

\(\alpha_\mu\)
&
\(0.1\)
&
\(0.1\)
\\

\(A\)
&
\(0.05\)
&
\(0.9\)
\\

\(\Delta\)
&
\(0.9\)
&
\(0.45\)
\\

\(p\)
&
\(2\)
&
\(12\)
\\

\(\beta_{\rm GX}\)
&
\(2\)
&
\(2\)
\\

\(M\)
&
\(3\times10^{-5}\)
&
\(1.4\times10^{-4}\)
\\

\(\phi_\star\)
&
\(5.3\)
&
\(4.9\)
\\

\hline

\(w_{\rm re}\)
&
\(0.05,\;0.20,\;0.30\)
&
\(0.94,\;0.96,\;0.98\)
\\

\end{tabular}
\end{ruledtabular}
\end{table*}

Both parameters affect reheating through the inflationary
background. Varying \(\lambda_{\rm GB}\) or $g_{_X}$ changes
\(H(N)\), \(\phi(N)\), and \(X(N)\), and therefore changes the
quantities \(H_\star\), \(c_{s\star}\), and \(\rho_{\rm end}\) that
enter the reheating matching relation. From
Eq.~\eqref{eq:Bmatch}, the corresponding shift is
\begin{equation}
\Delta f
=
\Delta\ln H_\star
-
\Delta\ln c_{s\star}
-
\frac{1}{4}\Delta\ln\rho_{\rm end}.
\label{eq:delta-f-scan}
\end{equation}
For a fixed value of \(w_{\rm re}\), this shift directly determines the change
in the reheating duration,
\begin{equation}
\Delta N_{\rm re}
=
\frac{4}{1-3w_{\rm re}}\Delta f .
\label{eq:delta-Nre-scan}
\end{equation}
The reheating temperature is affected by the same shift, but it also depends on the amount of dilution during reheating. Using Eq.~\eqref{eq:Tre-final}, one finds, for fixed \(w_{\rm re}\), 
\begin{equation}
\Delta\ln T_{\rm re}
=
\frac{1}{4}\Delta\ln\rho_{\rm end}
-
\frac{3}{4}(1+w_{\rm re})\Delta N_{\rm re}.
\label{eq:delta-Tre-scan}
\end{equation}
Accordingly, the change in the temperature comes from the change in the energy density at the end of inflation and change in the duration of reheating. A larger \(\rho_{\rm end}\) tends to increase \(T_{\rm re}\), while a longer reheating phase makes the exponential dilution stronger and therefore lowers
\(T_{\rm re}\).

The numerical values of $H_\star$, $c_{s\star}$, and
$\rho_{\rm end}$, which enter Eq.~\eqref{eq:Bmatch} and determine the shifts
in Eqs.~\eqref{eq:delta-f-scan}--\eqref{eq:delta-Tre-scan}, are reported in
Tables~\ref{tab:lambda-background-quantities}
and~\ref{tab:gx-background-quantities}.
The quantities \(H_\star\) and \(c_{s\star}\) are evaluated at pivot
exit, whereas \(\rho_{\rm end}\) is evaluated at the end of inflation.
These tables make it possible to trace the changes in the reheating
duration and temperature back to the modified inflationary
background.

\begin{table}[h]
\caption{
Inflationary background quantities entering the reheating matching
relations for the fixed-pivot \(\lambda_{\rm GB}\) scan. The Hubble
parameter \(H_\star\) and the scalar sound speed \(c_{s\star}\) are
evaluated at pivot exit, while \(\rho_{\rm end}\) is evaluated at the
end of inflation. All dimensionful quantities are given in reduced
Planck units.
}
\label{tab:lambda-background-quantities}
\centering
\begin{ruledtabular}
\begin{tabular}{cccc}
\(\lambda_{\rm GB}\)
&
\(H_\star\)
&
\(c_{s\star}\)
&
\(\rho_{\rm end}\)
\\
\hline
\(0\)
&
\(6.144\times10^{-6}\)
&
\(1.00000\)
&
\(2.715\times10^{-11}\)
\\
\(5\times10^{-4}\)
&
\(6.132\times10^{-6}\)
&
\(0.99851\)
&
\(2.705\times10^{-11}\)
\\
\(1\times10^{-3}\)
&
\(6.121\times10^{-6}\)
&
\(0.99703\)
&
\(2.694\times10^{-11}\)
\\
\(2\times10^{-3}\)
&
\(6.097\times10^{-6}\)
&
\(0.99410\)
&
\(2.673\times10^{-11}\)
\end{tabular}
\end{ruledtabular}
\end{table}

As shown in Table~\ref{tab:lambda-background-quantities},
the background quantities entering the matching relation change
only slightly over the selected $\lambda_{\rm GB}$ range.
Accordingly, increasing $\lambda_{\rm GB}$ from zero to
$2\times10^{-3}$ leads to only small changes in $f$,
$N_{\rm re}$, and $T_{\rm re}$.

\begin{table}[h]
\caption{
Inflationary background quantities entering the reheating matching
relations for the fixed-pivot $g_{_X}$ scan. The Hubble parameter
\(H_\star\) and the scalar sound speed \(c_{s\star}\) are evaluated at
pivot exit, while \(\rho_{\rm end}\) is evaluated at the end of
inflation. All dimensionful quantities are given in reduced Planck units.
}
\label{tab:gx-background-quantities}
\centering
\begin{ruledtabular}
\begin{tabular}{cccc}
$g_{_X}$
&
\(H_\star\)
&
\(c_{s\star}\)
&
\(\rho_{\rm end}\)
\\
\hline
\(0\)
&
\(1.553\times10^{-6}\)
&
\(0.97022\)
&
\(1.726\times10^{-12}\)
\\
\(0.005\)
&
\(1.508\times10^{-6}\)
&
\(0.96432\)
&
\(1.627\times10^{-12}\)
\\
\(0.010\)
&
\(1.462\times10^{-6}\)
&
\(0.96474\)
&
\(1.530\times10^{-12}\)
\\
\(0.020\)
&
\(1.370\times10^{-6}\)
&
\(0.97300\)
&
\(1.343\times10^{-12}\)
\\
\(0.050\)
&
\(1.078\times10^{-6}\)
&
\(0.98573\)
&
\(8.320\times10^{-13}\)
\end{tabular}
\end{ruledtabular}
\end{table}

Table~\ref{tab:gx-background-quantities} shows a larger variation
of the background quantities within the kinetic scan. As $g_{_X}$
increases from zero to $0.05$, both $H_\star$ and
$\rho_{\rm end}$ decrease. The sound speed varies
non-monotonically. It first decreases and then moves closer to
unity as $g_{_X}$ is increased.

The reduction in \(H_\star\) gives a negative contribution to
\(\Delta f\), whereas the decrease in \(\rho_{\rm end}\) contributes
with the opposite sign through
\(-\frac{1}{4}\Delta\ln\rho_{\rm end}\). For the selected parameter
range, the value of \(f\) decreases. Since the equation of state parameters
used in the kinetic scan satisfy \(w_{\rm re}>1/3\), the factor
\(4/(1-3w_{\rm re})\) is negative. The decrease in \(f\) therefore
produces a longer reheating stage. The resulting increase in
\(N_{\rm re}\), together with the lower value of
\(\rho_{\rm end}\), reduces \(T_{\rm re}\).

The corresponding reheating durations and temperatures are presented
in the following two tables. Within each scan, the values of
\(N_{\rm pivot}\), \(w_{\rm re}\), and the remaining model parameters
are kept fixed, so that the reported shifts are due only to the
variation of the selected scan parameter.

\begin{table*}[h]
\caption{
Variation of the reheating parameters in the
\(\lambda_{\rm GB}\)-scan for
\(w_{\rm re}=0.05\), \(0.20\), and \(0.30\).
For each fixed value of \(w_{\rm re}\), the shift
\(\Delta N_{\rm re}\) is measured relative to the corresponding
value at \(\lambda_{\rm GB}=0\).
}
\label{tab:lambdaGB-reheating-scan}
\begin{ruledtabular}
\begin{tabular}{ccccc}
\(\lambda_{\rm GB}\)
&
\(w_{\rm re}\)
&
\(N_{\rm re}\)
&
\(\Delta N_{\rm re}\)
&
\(T_{\rm re}\,[{\rm GeV}]\)
\\
\hline

\(0\)
& \(0.05\)
& \(3.439\)
& \(0.000\)
& \(1.522\times10^{14}\)
\\
\(0\)
& \(0.20\)
& \(7.308\)
& \(0.000\)
& \(3.178\times10^{12}\)
\\
\(0\)
& \(0.30\)
& \(29.231\)
& \(0.000\)
& \(9.569\times10^{2}\)
\\
\hline

\(5\times10^{-4}\)
& \(0.05\)
& \(3.441\)
& \(0.002\)
& \(1.518\times10^{14}\)
\\
\(5\times10^{-4}\)
& \(0.20\)
& \(7.313\)
& \(0.005\)
& \(3.160\times10^{12}\)
\\
\(5\times10^{-4}\)
& \(0.30\)
& \(29.252\)
& \(0.021\)
& \(9.366\times10^{2}\)
\\
\hline

\(10^{-3}\)
& \(0.05\)
& \(3.444\)
& \(0.005\)
& \(1.513\times10^{14}\)
\\
\(10^{-3}\)
& \(0.20\)
& \(7.318\)
& \(0.010\)
& \(3.142\times10^{12}\)
\\
\(10^{-3}\)
& \(0.30\)
& \(29.273\)
& \(0.042\)
& \(9.171\times10^{2}\)
\\
\hline

\(2\times10^{-3}\)
& \(0.05\)
& \(3.449\)
& \(0.010\)
& \(1.505\times10^{14}\)
\\
\(2\times10^{-3}\)
& \(0.20\)
& \(7.328\)
& \(0.020\)
& \(3.108\times10^{12}\)
\\
\(2\times10^{-3}\)
& \(0.30\)
& \(29.313\)
& \(0.082\)
& \(8.802\times10^{2}\)
\\

\end{tabular}
\end{ruledtabular}
\end{table*}

\begin{table*}[h]
\caption{
Variation of the reheating parameters in the \(g_X\)-scan.
For each fixed value of \(w_{\rm re}\), the shift
\(\Delta N_{\rm re}\) is measured relative to the corresponding
value at $g_{_X}=0$.
}
\label{tab:gx-reheating-scan}

\begin{ruledtabular}
\begin{tabular}{ccccc}

$g_{_X}$
&
\(w_{\rm re}\)
&
\(N_{\rm re}\)
&
\(\Delta N_{\rm re}\)
&
\(T_{\rm re}\,[{\rm GeV}]\)
\\
\hline

\(0\)     & \(0.94\) & \(26.210\) & \(0.000\) & \(3.142\times10^{-2}\) \\
\(0.005\) & \(0.94\) & \(26.229\) & \(0.019\) & \(3.012\times10^{-2}\) \\
\(0.010\) & \(0.94\) & \(26.264\) & \(0.054\) & \(2.820\times10^{-2}\) \\
\(0.020\) & \(0.94\) & \(26.354\) & \(0.144\) & \(2.392\times10^{-2}\) \\
\(0.050\) & \(0.94\) & \(26.646\) & \(0.436\) & \(1.389\times10^{-2}\) \\

\hline

\(0\)     & \(0.96\) & \(25.374\) & \(0.000\) & \(7.252\times10^{-2}\) \\
\(0.005\) & \(0.96\) & \(25.392\) & \(0.018\) & \(6.956\times10^{-2}\) \\
\(0.010\) & \(0.96\) & \(25.426\) & \(0.052\) & \(6.520\times10^{-2}\) \\
\(0.020\) & \(0.96\) & \(25.513\) & \(0.140\) & \(5.548\times10^{-2}\) \\
\(0.050\) & \(0.96\) & \(25.796\) & \(0.422\) & \(3.250\times10^{-2}\) \\

\hline

\(0\)     & \(0.98\) & \(24.589\) & \(0.000\) & \(1.590\times10^{-1}\) \\
\(0.005\) & \(0.98\) & \(24.607\) & \(0.018\) & \(1.525\times10^{-1}\) \\
\(0.010\) & \(0.98\) & \(24.639\) & \(0.050\) & \(1.431\times10^{-1}\) \\
\(0.020\) & \(0.98\) & \(24.724\) & \(0.135\) & \(1.221\times10^{-1}\) \\
\(0.050\) & \(0.98\) & \(24.998\) & \(0.409\) & \(7.218\times10^{-2}\) \\

\end{tabular}
\end{ruledtabular}
\end{table*}

As shown in Tables~\ref{tab:lambdaGB-reheating-scan} and
\ref{tab:gx-reheating-scan}, increasing either
\(\lambda_{\rm GB}\) or $g_{_X}$ increases \(N_{\rm re}\) and decreases
\(T_{\rm re}\) for each fixed value of \(w_{\rm re}\). In the \(\lambda_{\rm GB}\)-scan, these changes are small. The change in \(N_{\rm re}\) becomes larger when \(w_{\rm re}\) is closer to
\(1/3\). In the $g_{_X}$-scan, the same behavior is found for
\(w_{\rm re}=0.94\), \(0.96\), and \(0.98\).

In each scan, \(N_{\rm pivot}\), \(w_{\rm re}\), and the other model
parameters are kept fixed, while only \(\lambda_{\rm GB}\) or $g_{_X}$ is
changed. Since the two scans use different pivot separations and different model
parameters, their numerical shifts do not provide a direct comparison
of the effects of \(\lambda_{\rm GB}\) and \(g_X\).

To assess the benchmark dependence of the fixed-pivot trends, we
repeat both scans after changing parameters that control different
parts of the model. For the \(\lambda_{\rm GB}\) scan, we first increased
the field-dependence parameter from \(\alpha_\mu=0.1\) to
\(\alpha_\mu=0.5\). In a second test, we changed the E-model parameter
from \(\alpha=1\) to \(\alpha=2/3\). For the $g_{_X}$ scan, we first
considered the more strongly localized and more strongly saturated
choice $\alpha_\mu=0.5, \beta_{GX}=8,$
while keeping \(\alpha=1\) and \(\lambda_{\rm GB}=0.09\). We then
additionally changed the potential parameter and the overall
Gauss--Bonnet strength to $\alpha=\frac{2}{3}, \lambda_{\rm GB}=0.002.$

Rather than repeating the full set of reheating equation of state parameters,
we fix one representative value in each scan and focus on the changes
in \(N_{\rm re}\) and \(T_{\rm re}\). We use \(w_{\rm re}=0.30\) for
the \(\lambda_{\rm GB}\) scan and \(w_{\rm re}=0.98\) for the
$g_{_X}$ scan. Both values are already included in the reference
analysis.

\begin{table*}[h]
\caption{
Dependence of the fixed-pivot \(\lambda_{\rm GB}\) scan on the
benchmark choice for \(w_{\rm re}=0.30\). The table shows the values
at the two ends of the interval
\(0\leq\lambda_{\rm GB}\leq2\times10^{-3}\).
Intermediate scan points follow the same monotonic behavior.
}
\label{tab:lambda-alternative-benchmarks}
\centering
\begin{ruledtabular}
\begin{tabular}{cccccc}
Benchmark
&
\(N_{\rm re}(0)\)
&
\(N_{\rm re}(2\times10^{-3})\)
&
\(\Delta N_{\rm re}\)
&
\(T_{\rm re}(0)\,[{\rm GeV}]\)
&
\(T_{\rm re}(2\times10^{-3})\,[{\rm GeV}]\)
\\
\hline
\(\alpha=1,\ \alpha_\mu=0.1\)
&
\(29.231\)
&
\(29.313\)
&
\(+0.082\)
&
\(9.569\times10^{2}\)
&
\(8.802\times10^{2}\)
\\
\(\alpha=1,\ \alpha_\mu=0.5\)
&
\(29.231\)
&
\(29.271\)
&
\(+0.040\)
&
\(9.569\times10^{2}\)
&
\(9.189\times10^{2}\)
\\
\(\alpha=2/3,\ \alpha_\mu=0.1\)
&
\(23.722\)
&
\(23.730\)
&
\(+0.007\)
&
\(1.962\times10^{5}\)
&
\(1.938\times10^{5}\)
\end{tabular}
\end{ruledtabular}
\end{table*}

The first result is that the sign of the
\(\lambda_{\rm GB}\) trend is preserved in all three cases. Increasing
the overall Gauss--Bonnet strength slightly lengthens the effective
reheating stage and lowers its final temperature. However, the magnitude
of the effect is strongly benchmark dependent.

Increasing \(\alpha_\mu\) makes the factor
\(e^{-\alpha_\mu\phi}\) more strongly suppressed on the inflationary
plateau. Since the inflaton evolves from larger positive values of
\(\phi\) toward the minimum, the Gauss--Bonnet interaction becomes
concentrated in a shorter interval closer to the end of inflation.
The interaction is therefore not simply made uniformly weaker.
Instead, its overlap with most of the inflationary trajectory is
reduced, while its field dependence becomes more localized near the
end of inflation. Consequently, changing \(\lambda_{\rm GB}\) acts over a
shorter part of the trajectory. This reduces the shift in
\(N_{\rm re}\) from \(0.082\) to \(0.040\). Over the same range of $\lambda_{\rm GB}$,
$T_{\rm re}$ decreases from $9.569\times10^{2}\,{\rm GeV}$ to
$8.802\times10^{2}\,{\rm GeV}$ for $\alpha_\mu=0.1$, whereas for
$\alpha_\mu=0.5$ it decreases only to
$9.189\times10^{2}\,{\rm GeV}$.

The change from \(\alpha=1\) to \(\alpha=2/3\) has a different physical
meaning. For the E-model potential given in Eq.~\eqref{eq:potential},
this change increases \(b\) from \(\sqrt{2/3}\) to unity. The
transition from the inflationary plateau to the minimum is therefore
compressed into a narrower field interval. Near the minimum, $V(\phi)\simeq V_0b^2\phi^2$,
so that the curvature normalized by \(V_0\) increases by a factor
\(3/2\). The final evolution toward \(\epsilon_1=1\) is consequently
more strongly controlled by the potential itself.

This distinction is visible in Table~\ref{tab:lambda-alternative-benchmarks}.
Changing \(\alpha\) clearly shifts the baseline reheating
prediction even at \(\lambda_{\rm GB}=0\), reducing \(N_{\rm re}\)
from \(29.231\) to \(23.722\) and raising \(T_{\rm re}\). At the same
time, the additional variation considered by increasing
\(\lambda_{\rm GB}\) becomes very small:
\(\Delta N_{\rm re}\simeq0.007\). The narrower exit region leaves the
weak Gauss--Bonnet interaction less opportunity to alter the final
inflationary trajectory.

The $\lambda_{\rm GB}$ trend therefore retains the same sign
across the benchmarks considered here, but its magnitude is not uniform.
It becomes weaker when the coupling is more localized in field space
and becomes almost negligible when the end of inflation is more
strongly controlled by the curvature of the potential.

\begin{table*}[h]
\caption{
Dependence of the fixed-pivot \(g_{_X}\) scan on the benchmark choice
for \(w_{\rm re}=0.98\). The table shows the values at the two ends
of the interval \(0\leq g_{_X}\leq0.05\). Intermediate scan points
follow the indicated monotonic behavior.
}
\label{tab:gx-alternative-benchmarks}
\centering
\begin{ruledtabular}
\begin{tabular}{cccccc}
Benchmark
&
\(N_{\rm re}(0)\)
&
\(N_{\rm re}(0.05)\)
&
\(\Delta N_{\rm re}\)
&
\(T_{\rm re}(0)\,[{\rm GeV}]\)
&
\(T_{\rm re}(0.05)\,[{\rm GeV}]\)
\\
\hline
\(\alpha=1,\ \alpha_\mu=0.1,\ \beta_{GX}=2,\ 
\lambda_{\rm GB}=0.09\)
&
\(24.589\)
&
\(24.998\)
&
\(+0.409\)
&
\(1.590\times10^{-1}\)
&
\(7.218\times10^{-2}\)
\\
\(\alpha=1,\ \alpha_\mu=0.5,\ \beta_{GX}=8,\ 
\lambda_{\rm GB}=0.09\)
&
\(23.574\)
&
\(23.515\)
&
\(-0.058\)
&
\(1.121\)
&
\(1.201\)
\\
\(\alpha=2/3,\ \alpha_\mu=0.5,\ \beta_{GX}=8,\ 
\lambda_{\rm GB}=0.002\)
&
\(23.720\)
&
\(23.713\)
&
\(-0.0065\)
&
\(9.904\times10^{-1}\)
&
\(9.995\times10^{-1}\)
\end{tabular}
\end{ruledtabular}
\end{table*}

The $g_{_X}$ scan shows a qualitatively different behavior. In the reference benchmark, increasing $g_{_X}$ increases $N_{\rm re}$ and leads to a clear decrease in $T_{\rm re}$. This
occurs when the overall Gauss--Bonnet strength is relatively large,
\(\lambda_{\rm GB}=0.09\), and the kinetic gate has the saturation
parameter \(\beta_{GX}=2\). In this case, the explicit \(X\)
dependence remains sufficiently active over the inflationary
trajectory for \(g_X\) to modify the late-time background
clearly.

The choice $\alpha_\mu=0.5, \beta_{GX}=8$,
changes the physical role of the kinetic gate. The larger value of
\(\alpha_\mu\) concentrates the coupling closer to the end of
inflation, while the larger value of \(\beta_{GX}\) lowers the
maximum value of $G(X)=\frac{u}{1+\beta_{GX}u}$
from \(1/2\) to \(1/8\). It also suppresses the sensitivity of the
gate to changes in \(X\), since $G_{,X}\propto
\frac{1}{\left(1+\beta_{GX}u\right)^2}$.
The kinetic sector therefore changes from a relatively broad
modification of the trajectory to a more localized and more strongly
bounded correction.

Under this change, the trend reverses. Increasing $g_{_X}$ from zero
to $0.05$ decreases $N_{\rm re}$ from $23.574$ to $23.515$ and
increases $T_{\rm re}$ from $1.121\,{\rm GeV}$ to
$1.201\,{\rm GeV}$. The reversal shows that increasing $g_X$ does not lead to a uniform
change in the reheating duration across different parameter sets. It controls the amplitude of a bounded kinetic gate, and the effect of the gate depends on the region crossed by the
inflationary trajectory and on whether the gate remains sensitive
to changes in $X$ or has already entered the saturated regime. This behavior should not be described as a simple change in the effective friction of the inflaton. The phase-space coupling modifies both
the effective kinetic coefficient and the additional force terms in
the scalar equation. Their combined effect changes when the
field-space localization and kinetic saturation are varied.
Consequently, increasing $g_{_X}$ can produce opposite changes in
$N_{\rm re}$ for different parameter sets.

The third benchmark moves the model further toward a
potential-dominated regime. Reducing
\(\lambda_{\rm GB}\) from \(0.09\) to \(0.002\) weakens the overall
Gauss--Bonnet interaction. At the same time,
 for $\alpha=2/3$, the E-model potential approaches
its minimum over a narrower field interval. The explicit kinetic dependence is then multiplied by a much smaller overall Gauss--Bonnet strength, while the final inflationary
trajectory evolves more rapidly.

The small value $\Delta N_{\rm re}\simeq-0.0065$
shows that the $g_{_X}$ dependence has become almost inactive. The
negative sign found in the previous alternative benchmark remains,
but its magnitude is reduced by nearly one order of magnitude. This
is the approach to a decoupling regime. Changing $g_{_X}$ has little
effect because it modulates a Gauss--Bonnet sector that is already
weak, localized, and strongly saturated.

The two scans therefore show different types of benchmark
dependence. The effect of \(\lambda_{\rm GB}\) retains its sign but
can become strongly suppressed. By contrast, the effect of $g_{_X}$
can change sign and can eventually become negligible. This difference
is natural because \(\lambda_{\rm GB}\) rescales the complete
Gauss--Bonnet interaction, whereas $g_{_X}$ changes only its bounded
kinetic dependence. The latter is more sensitive to the trajectory,
the field-space localization, and the saturation properties of the
gate.

These benchmark comparisons show that the reheating trends are controlled by the
dynamical regime of the model rather than by the varied parameter
alone. A larger \(\lambda_{\rm GB}\) gives the same direction of
change in the benchmarks examined here, although the effect becomes
much weaker when the potential controls the end of inflation more
strongly. The $g_{_X}$ trend is less robust. It can be positive,
negative, or nearly absent depending on the field localization,
kinetic saturation, potential curvature, and overall Gauss--Bonnet
strength. The fixed-pivot results should therefore be interpreted as
benchmark-dependent predictions of the complete
\(\mu(\phi,X)\) model, rather than as universal monotonic relations
between a single varied parameter and the reheating duration.

\subsection{CMB and reheating constraints}
\label{subsec:cmb-and-reheating}

The representative and alternative parameter sets introduced in
Subsec.~\ref{subsec:fixed-pivot-parameter-scans} are now used to study the CMB and their reheating constraints.
We first compare their fixed-pivot predictions in the $(n_s,r)$ plane with CMB
constraints using the observational contours obtained from
the Planck 2018, BICEP/Keck 2018, and ACT DR6 data combinations
\cite{Planck2018Inflation,BKKeck2021,ACTDR6Louis2025}. We then evaluate their reheating constraints by varying
$N_{\rm pivot}$ for four representative values of the effective, or equivalently
e-fold-averaged, reheating equation-of-state parameter.

\begin{figure}[h]
    \centering
    \includegraphics[width=0.7\columnwidth]{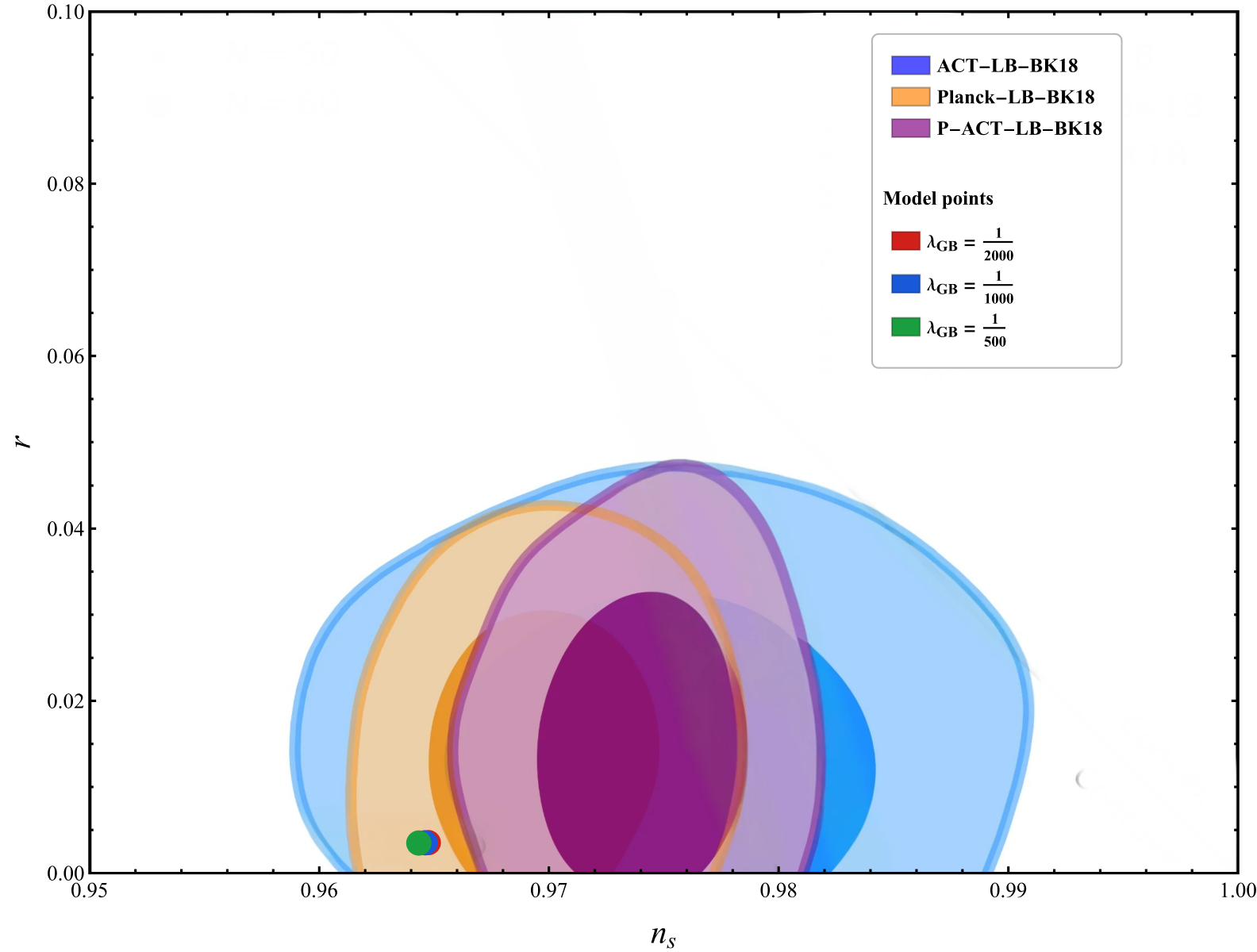}
    \caption{
    Predictions of the model in the \((n_s,r)\) plane for
    \(\lambda_{\rm GB}=1/2000\), \(1/1000\), and \(1/500\), with
    \(g_{_X}=10^{-3}\) and \(N_{\rm pivot}=55\).
    The darker and lighter shaded contours denote the joint \(68\%\) and
    \(95\%\) confidence regions, respectively, obtained from
    ACT--LB--BK18, Planck--LB--BK18, and the combined
    Planck--ACT--LB--BK18 data sets.
    }
    \label{fig:lambda-cmb-contour}
\end{figure}

For the representative $\lambda_{\rm GB}$ scan, we consider the three
nonzero values $\lambda_{\rm GB}
=\frac{1}{2000},
\frac{1}{1000},
\frac{1}{500}$,
with $g_{_X}=10^{-3}$. The remaining parameters are kept at the values
used in the fixed-pivot $\lambda_{\rm GB}$ scan, and the
inflationary observables are evaluated at $N_{\rm pivot}=55$.
The numerical solutions give 
\begin{align}
\lambda_{\rm GB}=\frac{1}{2000}:&
\qquad
n_s=0.964300,
\qquad
r=3.627\times10^{-3},
\nonumber\\
\lambda_{\rm GB}=\frac{1}{1000}:&
\qquad
n_s=0.964164,
\qquad
r=3.611\times10^{-3},
\nonumber\\
\lambda_{\rm GB}=\frac{1}{500}:&
\qquad
n_s=0.963886,
\qquad
r=3.579\times10^{-3}.
\label{eq:lambda-cmb-values}
\end{align}
These predictions are shown in
Fig.~\ref{fig:lambda-cmb-contour}.

The three predictions are closely located in the low-\(r\) region.
Increasing \(\lambda_{\rm GB}\) mainly shifts the result toward smaller
values of \(n_s\), while its effect on \(r\) remains small. All three points lie inside the displayed \(95\%\) confidence regions of the ACT--LB--BK18 and Planck--LB--BK18 data combinations, while remaining
outside the corresponding \(68\%\) regions. For the combined
P--ACT--LB--BK18 data set, the three predictions lie outside the displayed
\(95\%\) contour on its lower-\(n_s\) side. The point
with \(\lambda_{\rm GB}=1/2000\) is closest to the combined contour, whereas
\(\lambda_{\rm GB}=1/500\) gives the smallest value of the scalar spectral
index. The change in their positions is therefore mainly controlled by
\(n_s\), since the three values of \(r\) are already small.

\begin{figure}[h]
    \centering
    \includegraphics[width=0.7\columnwidth]
    {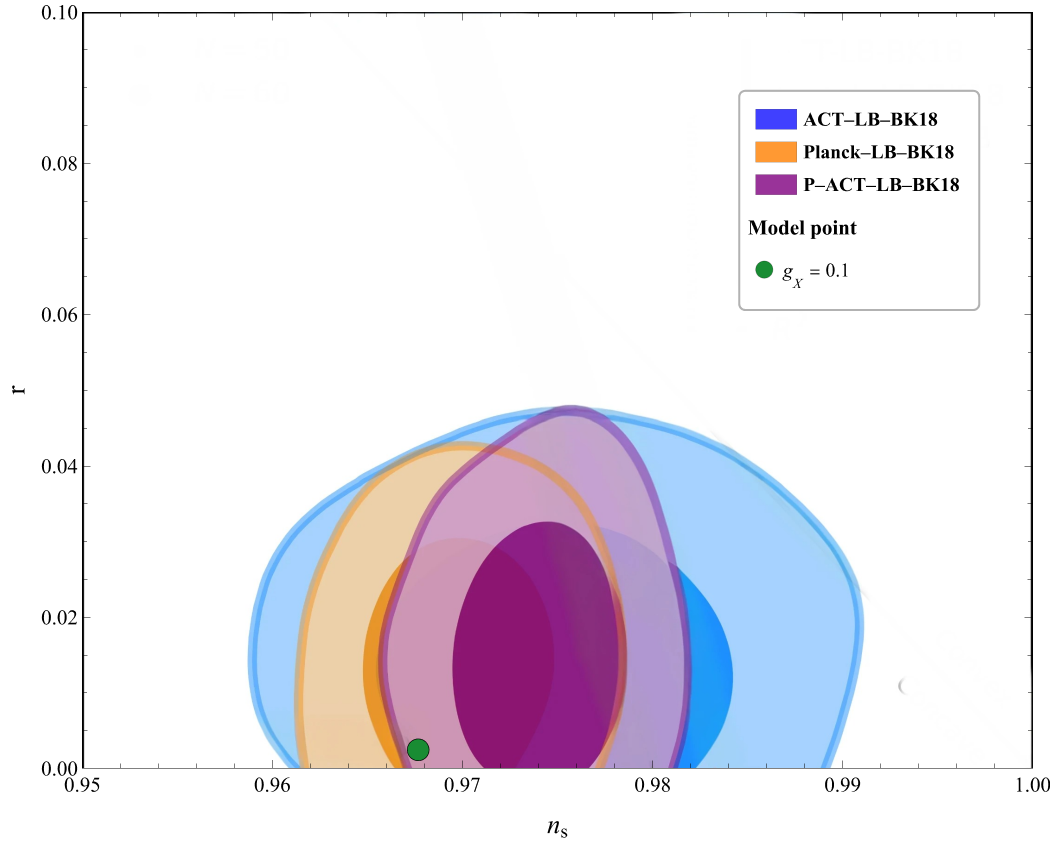}
    \caption{
    Prediction of the representative kinetic case in the \((n_s,r)\)
    plane for \(g_{_X}=0.1\), with \(\lambda_{\rm GB}=0.09\) and
    \(N_{\rm pivot}=67\). The darker and lighter shaded contours denote the
    joint \(68\%\) and \(95\%\) confidence regions, respectively, obtained
    from ACT--LB--BK18, Planck--LB--BK18, and the combined
    Planck--ACT--LB--BK18 data sets.
    }
    \label{fig:gx-cmb-contour}
\end{figure}

For the representative kinetic case, we consider $g_{_X}=0.1$ at
fixed $\lambda_{\rm GB}=0.09$. The inflationary observables are
evaluated at $N_{\rm pivot}=67$, while the remaining parameters are
kept at the values used in the corresponding fixed-pivot scan. The
numerical solution gives
\begin{equation}
n_s=0.967679,
\qquad
r=2.462\times10^{-3}.
\label{eq:gx-cmb-values}
\end{equation}
The corresponding point is shown in
Fig.~\ref{fig:gx-cmb-contour}.

The representative kinetic point lies inside the displayed $68\%$
confidence region of Planck--LB--BK18. It is also contained within
the displayed $95\%$ ACT--LB--BK18 confidence region, although it remains
outside the corresponding $68\%$ contour. For the combined
P--ACT--LB--BK18 data set, the point lies inside the displayed
$95\%$ confidence region and close to its lower-$n_s$ boundary.

\begin{figure*}[h]
    \centering
    \includegraphics[
        width=0.7\textwidth
    ]{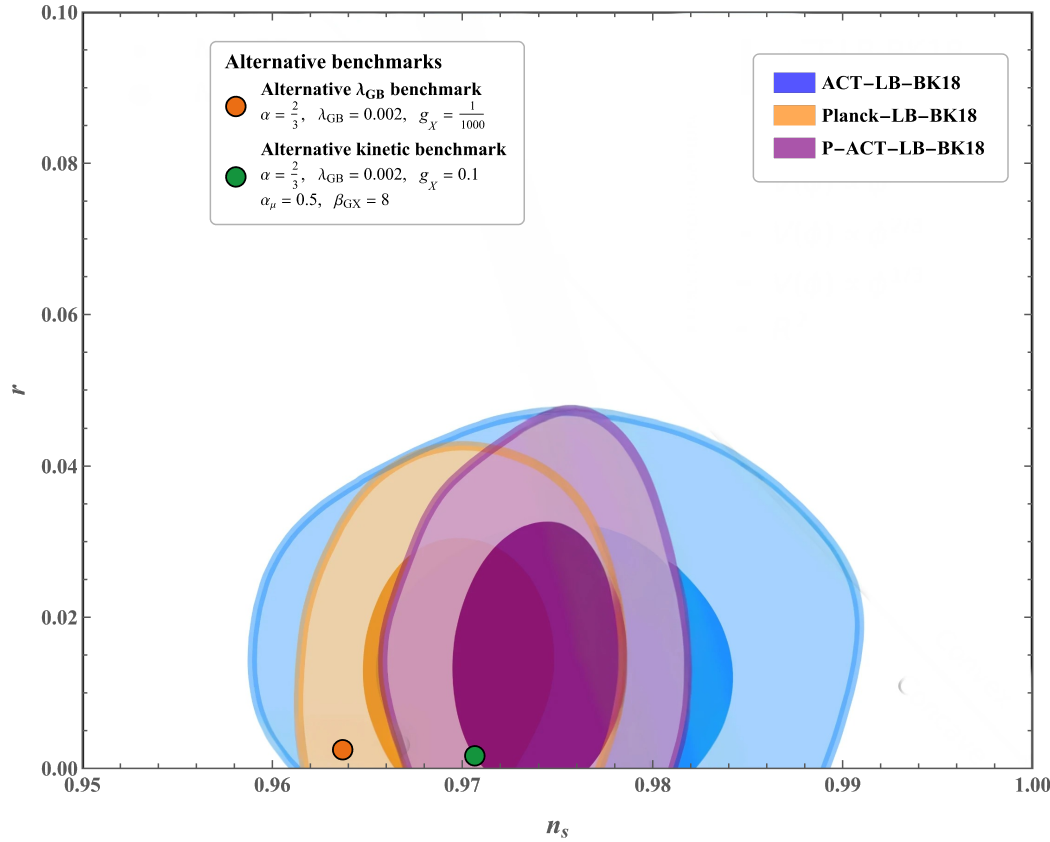}

    \caption{
    Predictions of the two alternative benchmarks used
    in the extended reheating analysis. The orange point denotes the
    alternative \(\lambda_{\rm GB}\) benchmark with
    $\alpha=2/3$, $\lambda_{\rm GB}=0.002$, and $g_{_X}=10^{-3}$.
    The green point denotes the alternative kinetic benchmark with
    $\alpha=2/3$, $\lambda_{\rm GB}=0.002$, $g_X=0.1$,
    $\alpha_\mu=0.5$, and $\beta_{GX}=8$.
    The darker and lighter shaded contours represent the joint
    $68\%$ and $95\%$ confidence regions, respectively, obtained from
    the ACT--LB--BK18, Planck--LB--BK18, and combined
    Planck--ACT--LB--BK18 data sets.
    }
    \label{fig:alternative-benchmark-contour}
\end{figure*}

Considering the alternative benchmarks introduced in Tables~\ref{tab:lambda-alternative-benchmarks} and
\ref{tab:gx-alternative-benchmarks}, we next examine their fixed-pivot predictions in
the $(n_s,r)$ plane. The
alternative \(\lambda_{\rm GB}\) benchmark corresponds to the final row of
Table~\ref{tab:lambda-alternative-benchmarks}, while the alternative kinetic benchmark corresponds to
the final row of Table~\ref{tab:gx-alternative-benchmarks}. At the fixed pivot separations used in
the corresponding scans, their predictions are
\begin{equation}
    n_s=0.963703,
\qquad
r=2.47284\times10^{-3},
\end{equation}
for the alternative \(\lambda_{\rm GB}\) benchmark, and
\begin{equation}
    n_s=0.970660,
\qquad
r=1.66766\times10^{-3},
\end{equation}
for the alternative kinetic benchmark. Their positions relative to
the CMB constraints are shown in Fig.~\ref{fig:alternative-benchmark-contour}.

The two alternative benchmarks remain in the low-$r$ region, but
they move in different directions relative to the corresponding
representative points. For the \(\lambda_{\rm GB}\) case, the change from
the representative to the alternative benchmark gives
\begin{equation}
    \Delta n_s=-4.61\times10^{-4},
\qquad
\Delta r=-1.138\times10^{-3}.
\end{equation}
For the kinetic case, the corresponding changes are
\begin{equation}
    \Delta n_s=-4.61\times10^{-4},
\qquad
\Delta r=-1.138\times10^{-3}.
\end{equation}

The alternative \(\lambda_{\rm GB}\) benchmark point lies inside the displayed
$95\%$ confidence regions of ACT--LB--BK18 and
Planck--LB--BK18, while remaining outside their corresponding
$68\%$ confidence regions. For the combined P--ACT--LB--BK18 data set, it lies
outside the displayed $95\%$ contour on its lower-$n_s$ side. The
alternative kinetic point lies inside the displayed $68\%$
confidence regions of both ACT--LB--BK18 and Planck--LB--BK18. It
is also contained within the displayed $95\%$ combined contour and
lies close to the lower-$n_s$ boundary of the corresponding $68\%$ confidence
region.
The fixed-pivot comparison therefore identifies four benchmark
predictions in the observational plane. The representative and
alternative \(\lambda_{\rm GB}\) points remain close to the lower-$n_s$
side of the displayed contours, whereas the kinetic benchmarks
lead to larger values of $n_s$. Among the four cases, the
alternative kinetic benchmark has the largest scalar spectral index
and lies inside the displayed $68\%$ regions of both the
ACT--LB--BK18 and Planck--LB--BK18 data combinations.

The subsequent reheating constraint analysis therefore includes
the representative \(\lambda_{\rm GB}\) benchmark with
$\lambda_{\rm GB}=10^{-3}$ and $g_{_X}=10^{-3}$, the representative
kinetic benchmark with $g_{_X}=0.1$ and $\lambda_{\rm GB}=0.09$, and
the two alternative benchmarks defined in the final rows of
Tables~\ref{tab:lambda-alternative-benchmarks} and
\ref{tab:gx-alternative-benchmarks}.

For each of the four benchmarks, all model parameters are kept
fixed while $N_{\rm pivot}$ is varied over the interval
\begin{equation}
    20\leq N_{\rm pivot}\leq75.
\end{equation}
The reheating duration and temperature are calculated from
Eqs.~\eqref{eq:Nre-final} and \eqref{eq:Tre-final}, respectively. We consider four representative values of the effective, or equivalently
e-fold-averaged, reheating equation-of-state parameter,
$
w_{\rm re}
=
-\frac{1}{3},
0,
\frac{2}{3},
1.$
The radiation-like value $w_{\rm re}=1/3$ is not included because
the matching relation for $N_{\rm re}$ becomes degenerate in this
case. Within each benchmark, the four reheating branches meet at an
instantaneous-reheating point defined by $N_{\rm re}=0$. The
position of this point depends on the complete inflationary
background and therefore differs among the four parameter sets.

For the representative \(\lambda_{\rm GB}\) benchmark, the common point is
located at
\begin{equation}
    n_s^{\rm inst}\simeq0.964624,
\qquad
\log_{10}
\left(
\frac{T_{\rm re}^{\rm inst}}{\rm GeV}
\right)
\simeq15.358.
\end{equation}
For the alternative \(\lambda_{\rm GB}\) benchmark, the corresponding
values are
\begin{equation}
n_s^{\rm inst}\simeq0.964076,
\qquad
\log_{10}
\left(
\frac{T_{\rm re}^{\rm inst}}{\rm GeV}
\right)
\simeq15.335.
\end{equation}
The alternative benchmark therefore leads to only a small
displacement of the instantaneous-reheating point. This is
consistent with the strongly suppressed $\lambda_{\rm GB}$
dependence found for the final benchmark in Table~\ref{tab:lambda-alternative-benchmarks}.

For the representative kinetic benchmark, the
instantaneous-reheating point is located at
\begin{equation}
n_s^{\rm inst}\simeq0.964119,
\qquad
\log_{10}
\left(
\frac{T_{\rm re}^{\rm inst}}{{\rm GeV}}
\right)
\simeq14.820.
\end{equation}
For the alternative kinetic benchmark, the corresponding values
are
\begin{equation}
    n_s^{\rm inst}\simeq0.96294,
\qquad
\log_{10}
\left(
\frac{T_{\rm re}^{\rm inst}}{{\rm GeV}}
\right)
\simeq15.3.
\end{equation}
The alternative kinetic benchmark therefore shifts the
instantaneous-reheating point toward smaller $n_s$ and increases
the corresponding temperature from approximately
$6.60\times10^{14}\,{\rm GeV}$ to
$2.0\times10^{15}\,{\rm GeV}$.

\begin{figure*}[h]
    \centering
    \includegraphics[width=0.485\textwidth]
    {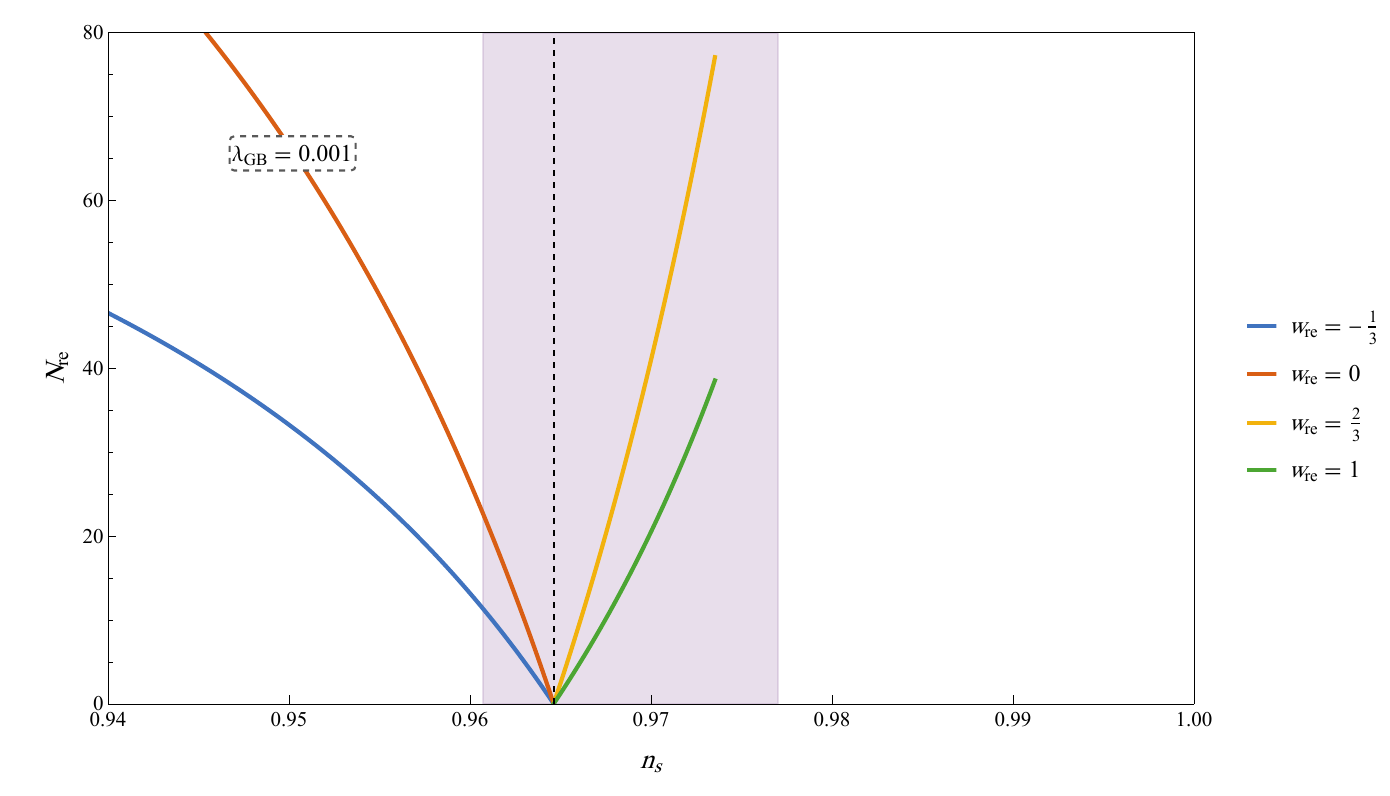}
    \hfill
    \includegraphics[width=0.485\textwidth]
    {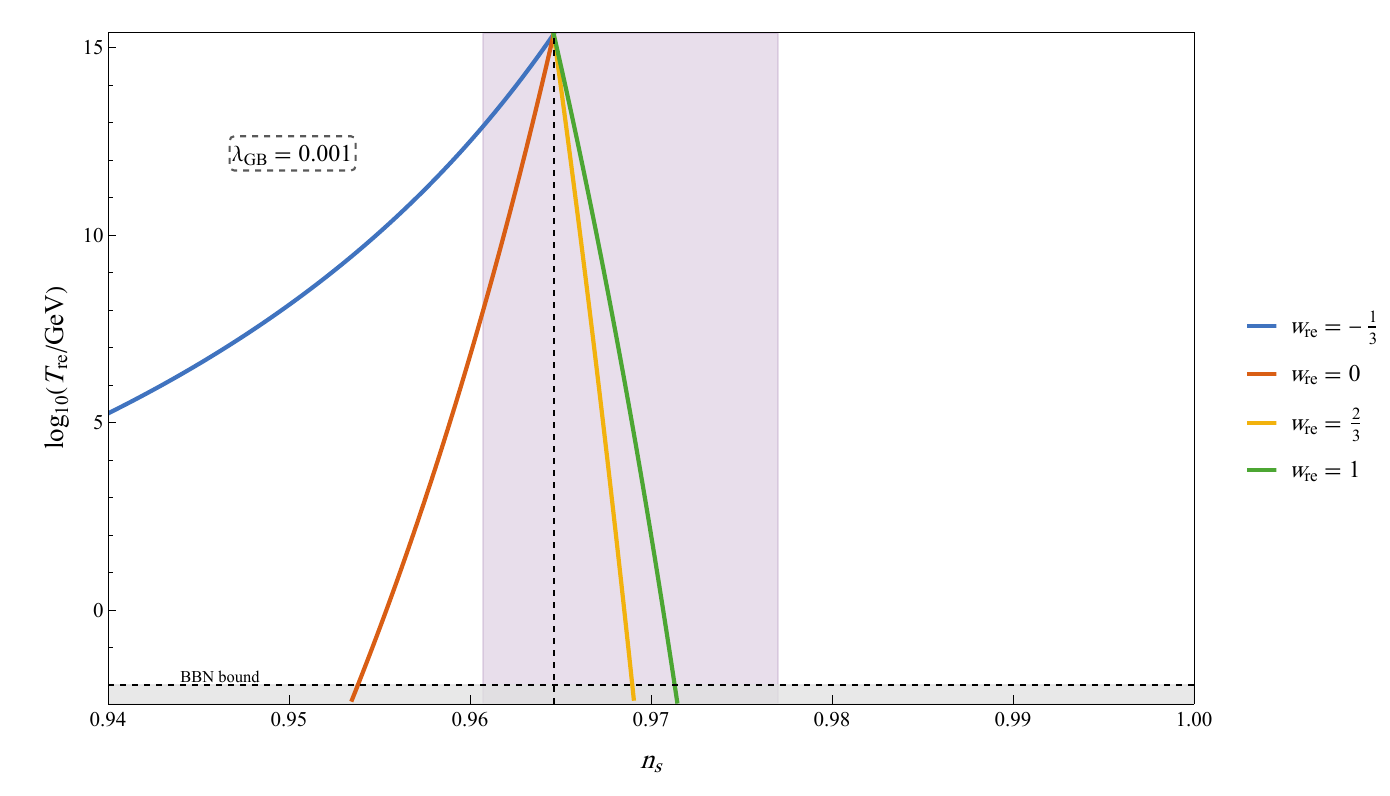}

    \caption{
    Reheating constraints for the representative $\lambda_{GB}$ benchmark
    \(\lambda_{\rm GB}=10^{-3}\) and \(g_{_X}=10^{-3}\).
    The left panel shows the reheating duration \(N_{\rm re}\) as a function
    of the scalar spectral index \(n_s\), while the right panel shows the
    corresponding reheating temperature \(T_{\rm re}\).
    The curves correspond to the four constant equations of state
    \(w_{\rm re}=-1/3\), \(0\), \(2/3\), and \(1\).
    The vertical shaded region indicates the adopted CMB range 
    of \(n_s\), and the vertical dashed line marks the common
    instantaneous-reheating point, where \(N_{\rm re}=0\).
    In the right panel, the horizontal dashed line denotes the approximate
    BBN bound \(T_{\rm re}=10\,{\rm MeV}\); the region below this limit is
    excluded.
    }
    \label{fig:lambda-reheating-curves}
\end{figure*}

\begin{figure}[h]
    \centering
    \includegraphics[width=0.485\textwidth]
    {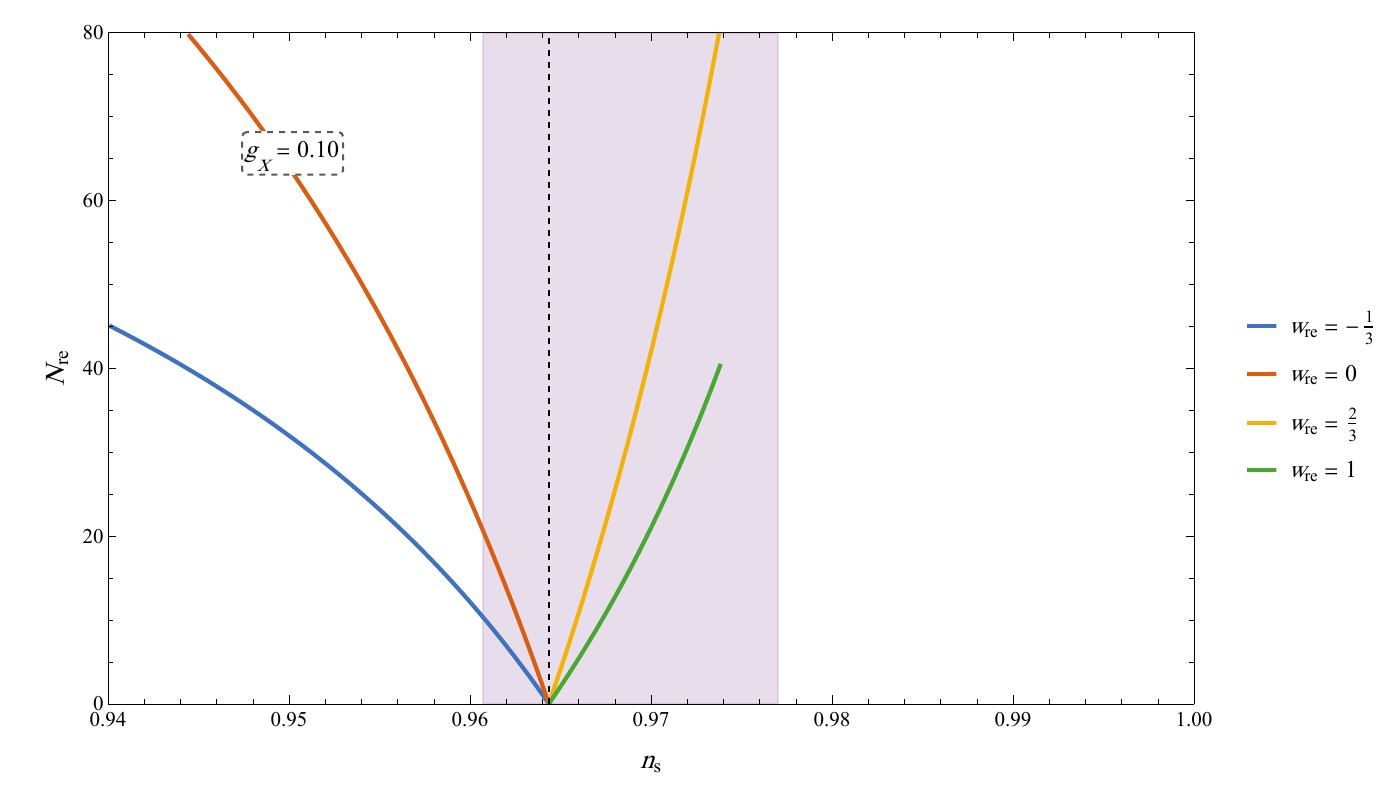}
    \hfill
    \includegraphics[width=0.485\textwidth]
    {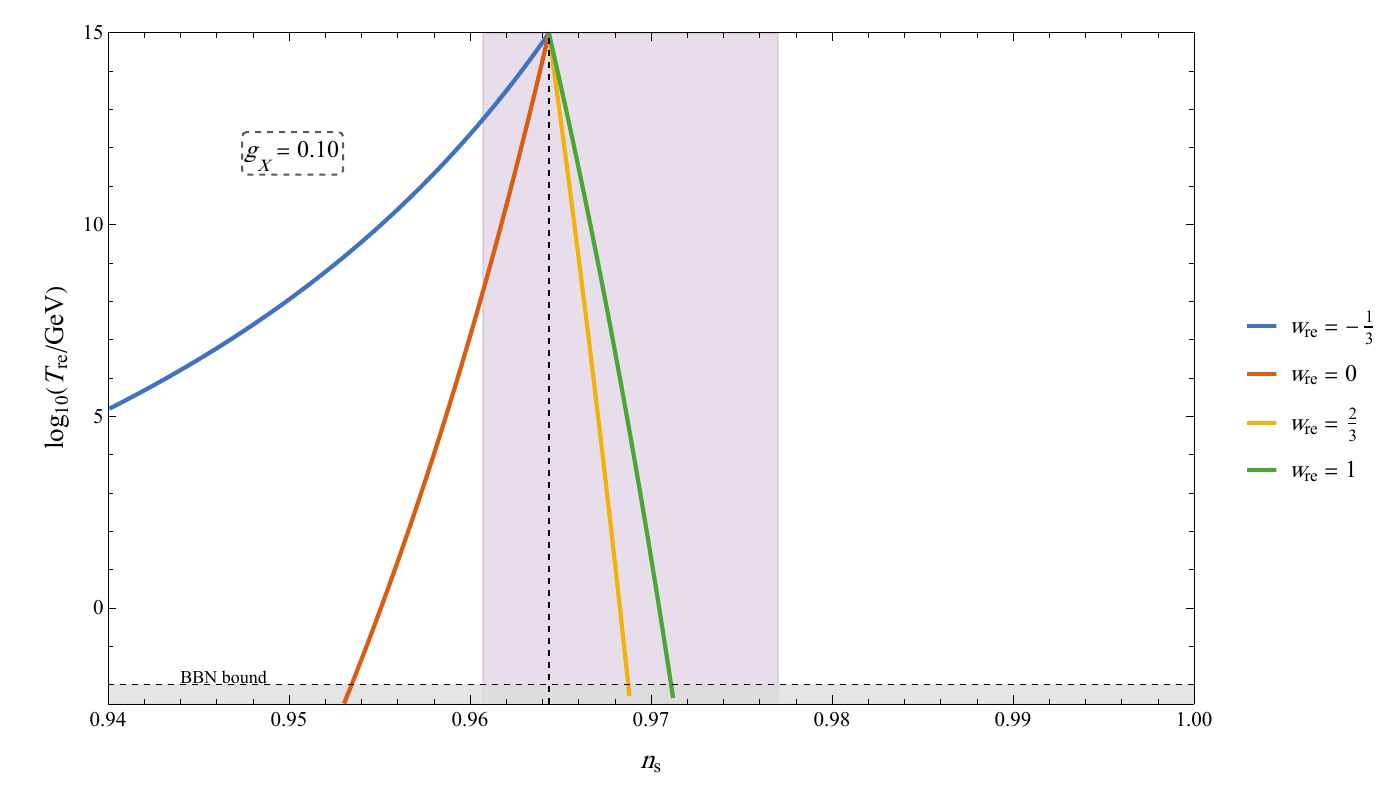}

    \caption{
    Reheating constraints for the representative kinetic benchmark
    \(g_{_X}=0.1\) and \(\lambda_{\rm GB}=0.09\).
    The left panel shows the reheating duration \(N_{\rm re}\), while the
    right panel shows the corresponding reheating temperature \(T_{\rm re}\),
    both as functions of the scalar spectral index \(n_s\).
    The four branches correspond to
    \(w_{\rm re}=-1/3\), \(0\), \(2/3\), and \(1\).
    The vertical shaded region indicates the adopted CMB range
    of \(n_s\), and the vertical dashed line marks the common
    instantaneous-reheating point. The horizontal dashed line in the right
    panel denotes the approximate BBN lower bound
    \(T_{\rm re}=10\,{\rm MeV}\).
    }
    \label{fig:gx-reheating-curves}
\end{figure}

\begin{figure*}[h]
    \centering
    \includegraphics[
        width=0.485\textwidth
    ]{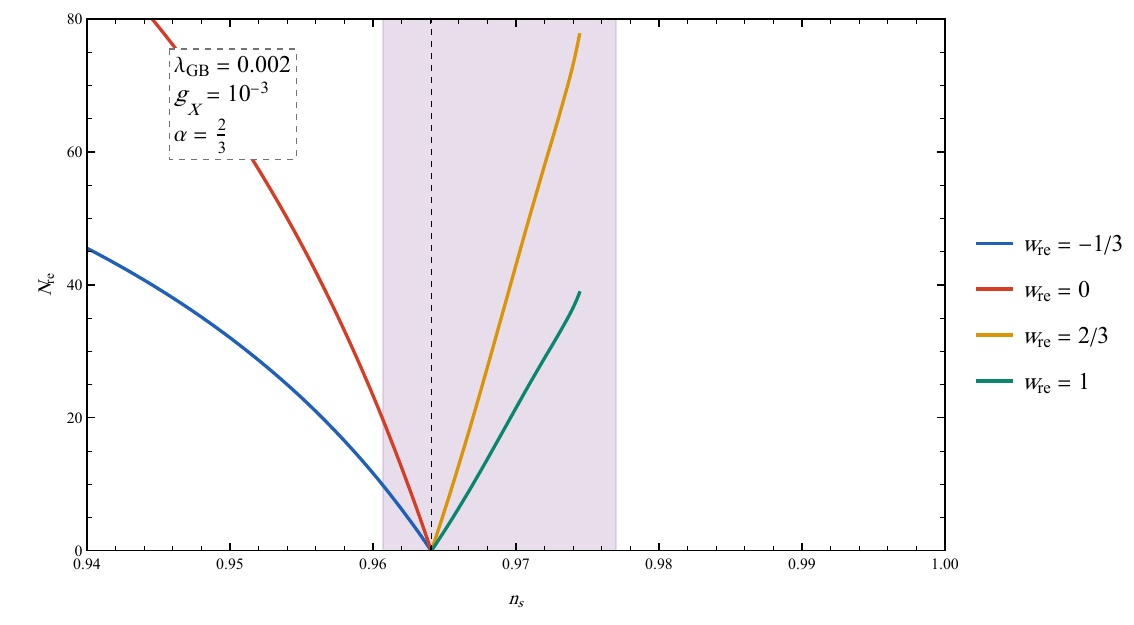}
    \hfill
    \includegraphics[
        width=0.485\textwidth
    ]{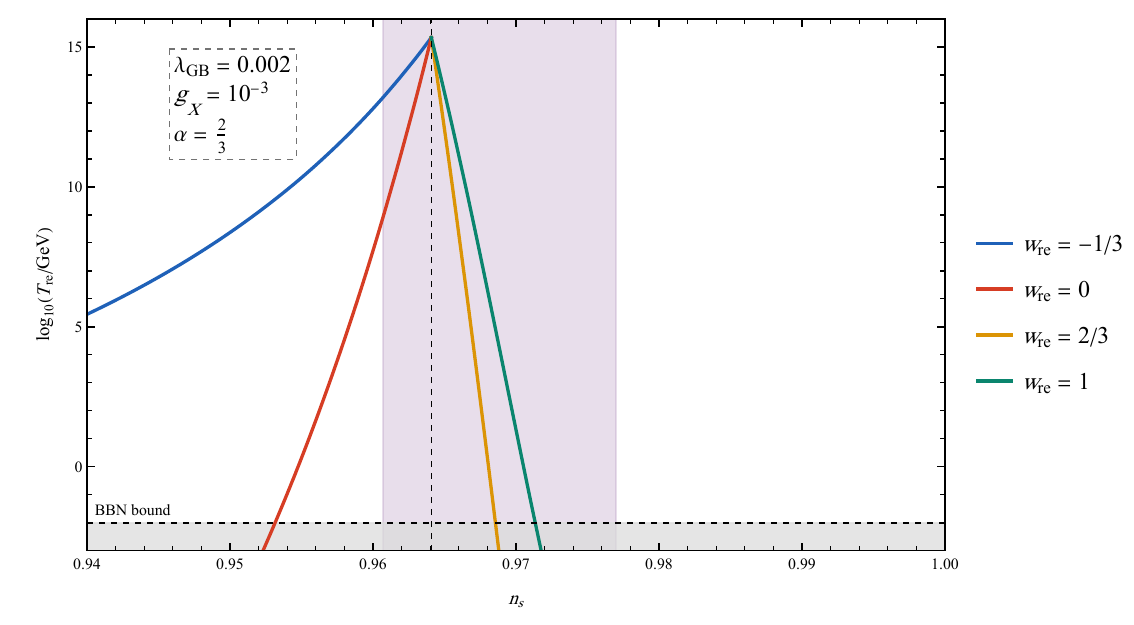}

    \caption{
    Reheating constraints for the alternative \(\lambda_{\rm GB}\) benchmark
    with $\alpha=2/3$, $\lambda_{\rm GB}=1/500$, and
    $g_{_X}=10^{-3}$. The left panel shows the reheating duration
    $N_{\rm re}$, while the right panel shows
    $\log_{10}(T_{\rm re}/{\rm GeV})$, both as functions of the scalar
    spectral index $n_s$. The four branches correspond to
    $w_{\rm re}=-1/3$, $0$, $2/3$, and $1$. The vertical shaded region
    indicates the adopted CMB range of $n_s$, and the
    vertical dashed line marks the common instantaneous-reheating
    point. In the right panel, the horizontal dashed line denotes the
    adopted BBN lower bound $T_{\rm re}=10\,{\rm MeV}$.
    }
    \label{fig:alternative-lambda-reheating}
\end{figure*}

\begin{figure*}[h]
    \centering
    \includegraphics[
        width=0.485\textwidth
    ]{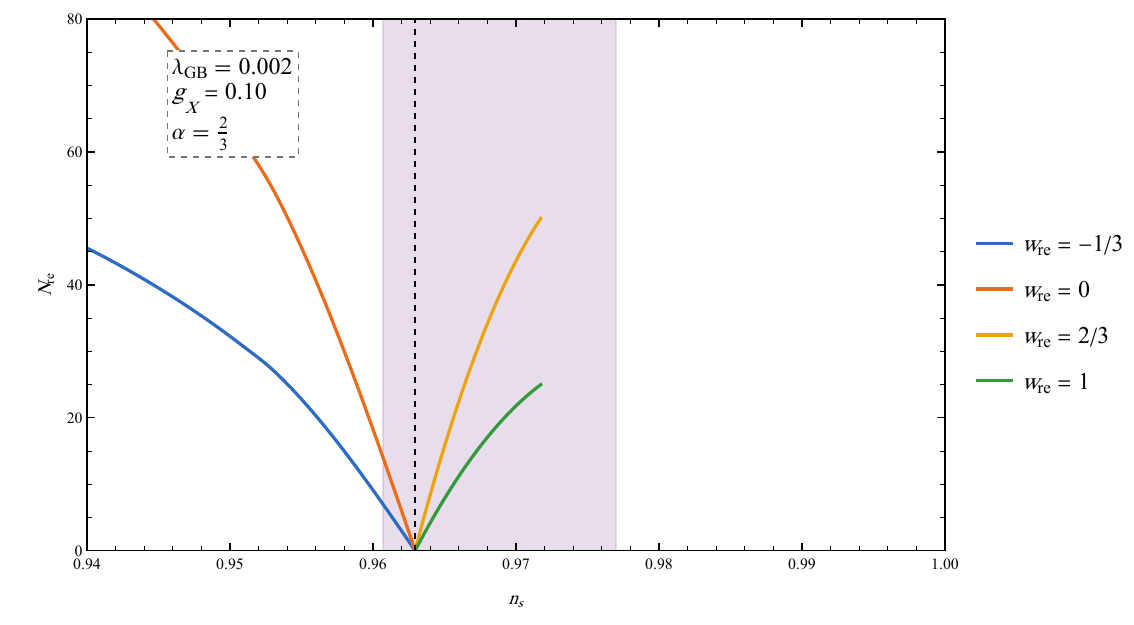}
    \hfill
    \includegraphics[
        width=0.485\textwidth
    ]{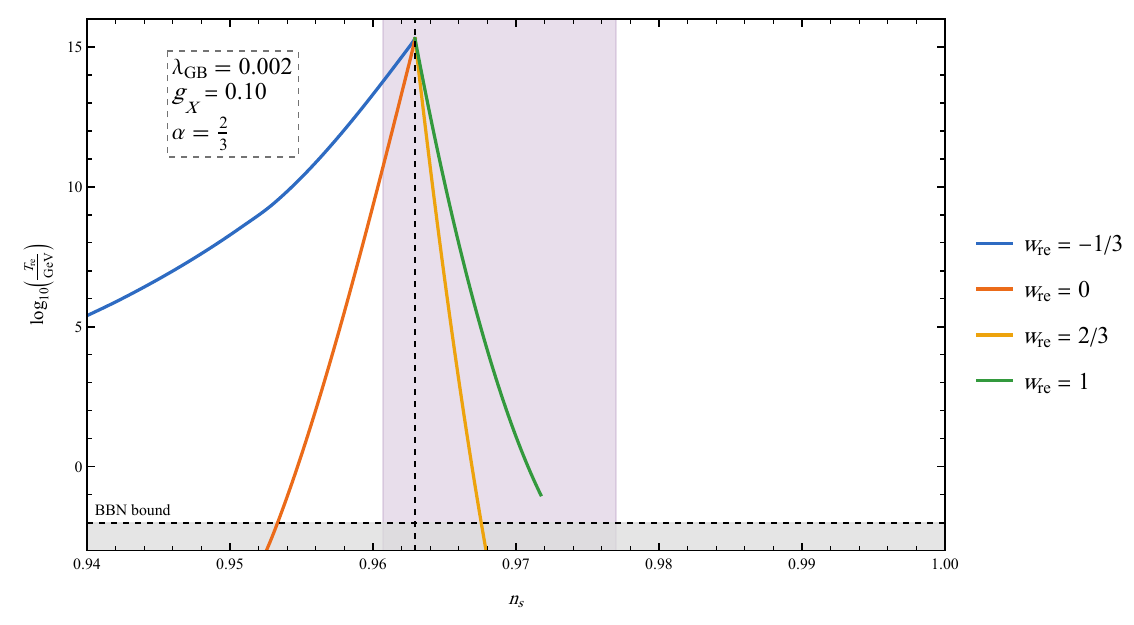}

    \caption{
    Reheating constraints for the alternative kinetic benchmark with
    $\alpha=2/3$, $\lambda_{\rm GB}=0.002$, $g_{_X}=0.1$,
    $\alpha_\mu=0.5$, and $\beta_{GX}=8$. The left panel shows the
    reheating duration $N_{\rm re}$, while the right panel shows
    $\log_{10}(T_{\rm re}/{\rm GeV})$, both as functions of the scalar
    spectral index $n_s$. The four branches correspond to
    $w_{\rm re}=-1/3$, $0$, $2/3$, and $1$. The vertical shaded region
    indicates the adopted CMB range of $n_s$, and the
    vertical dashed line marks the common instantaneous-reheating
    point. In the right panel, the horizontal dashed line denotes the
    adopted BBN lower bound $T_{\rm re}=10\,{\rm MeV}$.
    }
    \label{fig:alternative-kinetic-reheating}
\end{figure*}

We now compare the four benchmarks and their reheating constraints within the adopted CMB
comparison range
\begin{equation}
    0.9607\leq n_s\leq0.9770,
\end{equation}
which extends from the lower limit of the Planck $68\%$ confidence
interval to the upper limit of the ACT $68\%$ confidence
interval, as discussed in the CMB analysis. For each value of $w_{\rm re}$, we determine the corresponding ranges of $N_{\rm re}$ and $T_{\rm re}$ within this interval,
while requiring $ N_{\rm re}\geq0, T_{\rm re}\geq10\,{\rm MeV}.$

The reheating ranges obtained for the representative and alternative
$\lambda_{\rm GB}$ benchmarks are summarized in
Table~\ref{tab:lambda-reheating-ranges}.

\begin{table*}[h]
\centering
\caption{
Ranges of $N_{\rm re}$ and
$\log_{10}(T_{\rm re}/{\rm GeV})$ compatible with the adopted CMB
range and the BBN bound for the representative and alternative
$\lambda_{\rm GB}$ benchmarks.
}
\label{tab:lambda-reheating-ranges}
\begin{tabular}{c cc cc}
\hline\hline
& \multicolumn{2}{c}{Representative benchmark}
& \multicolumn{2}{c}{Alternative benchmark}
\\
\cline{2-3}\cline{4-5}
$w_{\rm re}$
&
$N_{\rm re}$
&
$\log_{10}(T_{\rm re}/{\rm GeV})$
&
$N_{\rm re}$
&
$\log_{10}(T_{\rm re}/{\rm GeV})$
\\
\hline
$-1/3$
&
$0$--$11.38$
&
$12.89$--$15.358$
&
$0$--$9.85$
&
$13.20$--$15.335$
\\

$0$
&
$0$--$22.76$
&
$7.94$--$15.358$
&
$0$--$19.70$
&
$8.92$--$15.335$
\\

$2/3$
&
$0$--$31.97$
&
$-2$--$15.358$
&
$0$--$31.93$
&
$-2$--$15.335$
\\

$1$
&
$0$--$26.65$
&
$-2$--$15.358$
&
$0$--$26.61$
&
$-2$--$15.335$
\\
\hline\hline
\end{tabular}
\end{table*}

Within the adopted CMB range, the representative and alternative
$\lambda_{\rm GB}$ benchmarks give similar reheating constraints.
For $w_{\rm re}=-1/3$ and $w_{\rm re}=0$, the alternative benchmark
allows shorter reheating durations and higher temperatures than the
representative case. For $w_{\rm re}=2/3$ and $w_{\rm re}=1$, the
allowed ranges of $N_{\rm re}$ and $T_{\rm re}$ are nearly the same
in the two benchmarks. The change of the $\lambda_{\rm GB}$
benchmark therefore has only a limited effect on the reheating
ranges compatible with the CMB and BBN constraints.

The corresponding results for the representative and alternative
kinetic benchmarks are summarized in
Table~\ref{tab:kinetic-reheating-ranges}.

\begin{table*}[h]
\centering
\caption{
Ranges of $N_{\rm re}$ and
$\log_{10}(T_{\rm re}/{\rm GeV})$ compatible with the adopted CMB
range and the BBN bound for the representative and alternative
kinetic benchmarks.
}
\label{tab:kinetic-reheating-ranges}
\begin{tabular}{c cc cc}
\hline\hline
& \multicolumn{2}{c}{Representative benchmark}
& \multicolumn{2}{c}{Alternative benchmark}
\\
\cline{2-3}\cline{4-5}
$w_{\rm re}$
&
$N_{\rm re}$
&
$\log_{10}(T_{\rm re}/{\rm GeV})$
&
$N_{\rm re}$
&
$\log_{10}(T_{\rm re}/{\rm GeV})$
\\
\hline
$-1/3$
&
$0$--$9.63$
&
$12.73$--$14.820$
&
$0$--$7.02$
&
$13.77$--$15.3$
\\

$0$
&
$0$--$19.26$
&
$8.55$--$14.820$
&
$0$--$14.05$
&
$10.72$--$15.3$
\\

$2/3$
&
$0$--$30.98$
&
$-2$--$14.820$
&
$0$--$31.87$
&
$-2$--$15.3$
\\

$1$
&
$0$--$25.82$
&
$-2$--$14.820$
&
$0$--$25$
&
$-1$--$15.3$
\\
\hline\hline
\end{tabular}
\end{table*}

Within the adopted CMB range, the representative and alternative
kinetic benchmarks show a clearer difference. For
$w_{\rm re}=-1/3$ and $w_{\rm re}=0$, the alternative benchmark
gives shorter reheating durations and higher temperatures. The
differences are smaller for $w_{\rm re}=2/3$ and $w_{\rm re}=1$,
although the allowed parts of these branches are more strongly
restricted by the BBN temperature bound. The reheating ranges are
therefore more sensitive to the kinetic benchmark than to the
$\lambda_{\rm GB}$ benchmark.

The two comparisons show that all four values of $w_{\rm re}$ retain
allowed ranges within the adopted CMB interval. The
$\lambda_{\rm GB}$ benchmarks remain close to each other, whereas
the kinetic benchmarks lead to a clearer change in the allowed
reheating duration and temperature. In both comparisons, the
$w_{\rm re}=2/3$ branches are the most strongly restricted by the
BBN bound.

\section{Varying Equation-of-State Parameter and Related Issues}
\label{subsec:varying-eos}

In the preceding analysis, the reheating stage was described in terms of a
constant effective equation-of-state parameter \(w_{\rm re}\). This assumption
provides a simple and model-independent connection between the inflationary
background and the subsequent thermal history
\cite{MunozKamionkowski2015,SahaAnandSriramkumar2020}. Nevertheless, the effective
equation of state is generally expected to evolve during reheating. Immediately
after inflation, the energy density may be dominated by a coherently oscillating
inflaton, while the production, interaction, and gradual thermalization of
relativistic particles continuously modify the composition of the cosmic fluid
\cite{KofmanLindeStarobinsky1997,LozanovAmin2017,SahaAnandSriramkumar2020}.
The effective equation-of-state parameter can therefore evolve from an
inflaton-dominated value toward the radiation value \(w=1/3\).

This generalization does not require a modification of the inflationary
background analysis performed in the preceding subsections. In particular, the
quantities \(H_\star\), \(c_{s\star}\), and \(\rho_{\rm end}\), obtained by
solving the phase-space-dependent Gauss--Bonnet background equations, remain
unchanged. The same values reported in Tables~II and III continue to provide
the inflationary input for the reheating matching relation. Only the dilution
of the energy density between the end of inflation and the completion of
reheating must be generalized.

Accordingly, the constant parameter \(w_{\rm re}\) can be replaced by a
time-dependent function \(w_{\rm re}(N)\), where \(N\) is now measured from the
end of inflation. The reheating continuity equation then becomes
\begin{equation}
\dot{\rho}_{\rm re}
+3H\left[1+w_{\rm re}(N)\right]\rho_{\rm re}=0.
\label{eq:varying_continuity_time}
\end{equation}
Using \(dN=Hdt\), this equation can be written as
\begin{equation}
\frac{d\rho_{\rm re}}{dN}
+3\left[1+w_{\rm re}(N)\right]\rho_{\rm re}=0.
\label{eq:varying_continuity_efold}
\end{equation}
Integrating from the end of inflation, \(N=0\), to the end of reheating,
\(N=N_{\rm re}\), gives
\begin{equation}
\rho_{\rm re}
=
\rho_{\rm end}
\exp\left[
-3\int_{0}^{N_{\rm re}}
\left(1+w_{\rm re}(N)\right)dN
\right],
\label{eq:varying_rho_general}
\end{equation}
or, equivalently,
\begin{equation}
\rho_{\rm re}
=
\rho_{\rm end}
\exp\left[
-3N_{\rm re}
-3\int_{0}^{N_{\rm re}}w_{\rm re}(N)\,dN
\right].
\label{eq:varying_rho_expanded}
\end{equation}
Equation~\eqref{eq:varying_rho_expanded} is the direct generalization of
Eq.~(32).

To relate this generalized description to the numerical analysis already
presented, we define the e-fold-averaged reheating equation-of-state parameter
as \cite{SahaAnandSriramkumar2020,GarciaPierre2023}
\begin{equation}
\overline{w}_{\rm re}
\equiv
\frac{1}{N_{\rm re}}
\int_{0}^{N_{\rm re}}w_{\rm re}(N)\,dN.
\label{eq:average_wre}
\end{equation}
The energy density at the completion of reheating can then be written as
\begin{equation}
\rho_{\rm re}
=
\rho_{\rm end}
\exp\left[
-3\left(1+\overline{w}_{\rm re}\right)N_{\rm re}
\right].
\label{eq:rho_average_wre}
\end{equation}
Therefore, the thermal-history matching relations derived for a constant
equation-of-state parameter retain exactly the same mathematical form after
the replacement
\begin{equation}
w_{\rm re}\longrightarrow\overline{w}_{\rm re}.
\label{eq:w_to_average_w}
\end{equation}
In particular, the generalized matching relation is
\begin{equation}
N_{\rm re}
-3\int_{0}^{N_{\rm re}}w_{\rm re}(N)\,dN
=4f,
\label{eq:general_matching_integral}
\end{equation}
where \(f\) is defined in Eq.~(42). In terms of the averaged parameter, this
relation reduces to
\begin{equation}
N_{\rm re}
=
\frac{4f}{1-3\overline{w}_{\rm re}}.
\label{eq:Nre_average_w}
\end{equation}
The corresponding reheating temperature becomes
\begin{equation}
T_{\rm re}
=
\left(
\frac{30\rho_{\rm end}}{\pi^2g_{\rm re}}
\right)^{1/4}
\exp\left[
-\frac{3}{4}
\int_{0}^{N_{\rm re}}
\left(1+w_{\rm re}(N)\right)dN
\right],
\label{eq:Tre_varying_general}
\end{equation}
which can be expressed as
\begin{equation}
T_{\rm re}
=
\left(
\frac{30\rho_{\rm end}}{\pi^2g_{\rm re}}
\right)^{1/4}
\exp\left[
-\frac{3}{4}
\left(1+\overline{w}_{\rm re}\right)N_{\rm re}
\right].
\label{eq:Tre_average_w}
\end{equation}

From the numerical point of view, this result allows the entire framework
presented in the preceding figures and tables to be preserved. The values of
\(w_{\rm re}\) adopted in Tables~I, IV, and V, as well as in the benchmark
comparisons of Tables~VI and VII, can equivalently be interpreted as selected
values of the averaged parameter \(\overline{w}_{\rm re}\). Similarly, the
constant values \(w_{\rm re}=-1/3,0,2/3,\) and \(1\), used to construct the
reheating curves in Figs.~4--7 and the allowed ranges in Tables~VIII and IX,
may be regarded as representative values of
\(\overline{w}_{\rm re}\) rather than necessarily as instantaneous values
maintained throughout the entire reheating stage.

It is useful to clarify explicitly how this generalization affects the
results obtained in Sec.~\ref{sec:numerical-results}. The thermal-history
matching relations do not depend separately on the instantaneous evolution of
$w_{\rm re}(N)$, but only on its e-fold average,
\begin{equation}
\overline{w}_{\rm re}
=
\frac{1}{N_{\rm re}}
\int_{0}^{N_{\rm re}}w_{\rm re}(N)\,dN.
\label{eq:same_average_condition}
\end{equation}
Consequently, the constant values of $w_{\rm re}$ employed in
Sec.~\ref{sec:numerical-results} should more generally be interpreted as
values of $\overline{w}_{\rm re}$. Any time-dependent reheating history having
the same e-fold average produces the same $N_{\rm re}$ and $T_{\rm re}$ within
the effective matching framework adopted here. The effect of a varying
reheating equation of state is therefore to change the mapping between a
microscopic reheating history and the effective parameter used in
Sec.~\ref{sec:numerical-results}, rather than to introduce an independent
correction to the matching relations. Histories with different time dependence
but identical $\overline{w}_{\rm re}$ are degenerate at this level of analysis,
whereas histories with different e-fold averages correspond to different
reheating branches and hence to different values of $N_{\rm re}$ and
$T_{\rm re}$. Thus, no recalculation of the inflationary background or of the
numerical quantities reported in the preceding tables is required for histories
with the same e-fold average~\cite{SahaAnandSriramkumar2020,GarciaPierre2023}.

The sensitivity of the reheating duration to the averaged equation of state
follows directly from Eq.~\eqref{eq:Nre_average_w},
\begin{equation}
\frac{\partial N_{\rm re}}{\partial \overline{w}_{\rm re}}
=
\frac{12f}{\left(1-3\overline{w}_{\rm re}\right)^2}.
\label{eq:Nre_average_sensitivity}
\end{equation}
This also explains why the matching becomes especially sensitive as
$\overline{w}_{\rm re}\rightarrow 1/3$.

A phenomenologically motivated description can be obtained by noting that
reheating involves a transition from an inflaton-dominated phase to radiation
domination. Near the minimum of the E-model potential,
$V(\phi)\propto\phi^2$, so that coherent inflaton oscillations have an
approximately matter-like averaged equation of state,
$\langle w_{\phi}\rangle\simeq0$. As radiation is progressively produced, the
total equation of state evolves toward $1/3$. For a mixture of the oscillating
inflaton and radiation one has
\begin{equation}
w_{\rm tot}(N)
=
\frac{w_{\phi}\rho_{\phi}(N)+\rho_r(N)/3}
{\rho_{\phi}(N)+\rho_r(N)}.
\label{eq:motivated_total_eos}
\end{equation}
This motivates a smooth phenomenological interpolation such as
\begin{equation}
w_{\rm re}(N)
=
w_i+
\frac{w_f-w_i}{1+\exp[-(N-N_t)/\Delta N]},
\label{eq:logistic_wre}
\end{equation}
where $w_i\simeq0$ describes the oscillating inflaton-dominated regime,
$w_f\simeq1/3$ corresponds to the approach to radiation domination, $N_t$
specifies the characteristic transition time, and $\Delta N$ controls the
duration of the transition.

As an alternative illustrative profile, a smooth approach to radiation domination can be represented
by

\begin{equation}
w_{\rm re}(N)
=
w_f-
\left(w_f-w_i\right)e^{-N/N_t},
\label{eq:exponential_wre}
\end{equation}
where \(N_t\) controls the characteristic transition timescale. In this case,
\begin{equation}
\int_{0}^{N_{\rm re}}w_{\rm re}(N)\,dN
=
w_fN_{\rm re}
-
(w_f-w_i)N_t
\left(
1-e^{-N_{\rm re}/N_t}
\right).
\label{eq:exponential_w_integral}
\end{equation}
Substitution into Eq.~\eqref{eq:general_matching_integral} gives
\begin{equation}
(1-3w_f)N_{\rm re}
+
3(w_f-w_i)N_t
\left(
1-e^{-N_{\rm re}/N_t}
\right)
=4f.
\label{eq:Nre_exponential_w}
\end{equation}

For this illustrative exponential profile, the relation is generally implicit and must be solved numerically for \(N_{\rm re}\). Once \(N_{\rm re}\) is determined, the
reheating temperature follows from
\begin{align}
T_{\rm re}
={}&
\left(
\frac{30\rho_{\rm end}}{\pi^2g_{\rm re}}
\right)^{1/4}
\exp\Bigg\{
-\frac{3}{4}
\Big[
(1+w_f)N_{\rm re}
\nonumber\\
&\hspace{3.5cm}
-
(w_f-w_i)N_t
\left(
1-e^{-N_{\rm re}/N_t}
\right)
\Big]
\Bigg\}.
\label{eq:Tre_exponential_w}
\end{align}
This parametrization provides a smooth phenomenological description of the
transition toward radiation domination while preserving the thermal-history
framework used throughout the present work.

A more dynamical determination of \(w_{\rm re}(N)\) would require extending the
modified background equations beyond the end-of-inflation condition
\(\epsilon_1=1\). The instantaneous equation-of-state parameter associated
with the phase-space-dependent Gauss--Bonnet sector can be defined as
\begin{equation}
w_{\phi}(N)
=
\frac{p_{\rm eff}(N)}{\rho_{\rm eff}(N)},
\label{eq:inflaton_eos}
\end{equation}
where \(\rho_{\rm eff}\) and \(p_{\rm eff}\) are given by Eqs.~(22) and (28),
respectively. The corresponding average is
\begin{equation}
\overline{w}_{\rm re}
=
\frac{1}{N_{\rm re}}
\int_{0}^{N_{\rm re}}
\frac{p_{\rm eff}(N)}{\rho_{\rm eff}(N)}\,dN.
\label{eq:dynamical_average_w}
\end{equation}
Such an inflaton-dominated evolution would not, however, fully describe the
production and thermalization of radiation. A complete dynamical treatment
would require a radiation component and an energy-transfer term \(Q\)
\cite{KofmanLindeStarobinsky1997,GarciaPierre2023,GhoshGhosh2025},
\begin{align}
\dot{\rho}_{\phi}
+3H\left(\rho_{\phi}+p_{\phi}\right)
&=-Q,
\label{eq:inflaton_energy_transfer}
\\
\dot{\rho}_{r}+4H\rho_r
&=Q.
\label{eq:radiation_energy_transfer}
\end{align}
For perturbative decay, a conventional phenomenological choice is
\cite{GhoshGhosh2025}
\begin{equation}
Q=\Gamma_{\phi}\dot{\phi}^{\,2},
\label{eq:perturbative_Q}
\end{equation}
although the precise form of \(Q\) should be derived consistently from the
modified scalar-field energy balance in the present phase-space-dependent
Gauss--Bonnet model.

The total equation-of-state parameter of the coupled inflaton--radiation system
would then be
\begin{equation}
w_{\rm tot}(N)
=
\frac{
p_{\phi}(N)+\rho_r(N)/3
}{
\rho_{\phi}(N)+\rho_r(N)
}.
\label{eq:total_reheating_eos}
\end{equation}
As radiation becomes dominant, \(w_{\rm tot}\) approaches \(1/3\). The end of
reheating may then be defined dynamically through
\begin{equation}
\rho_r(N_{\rm re})=\rho_{\phi}(N_{\rm re}),
\label{eq:reheating_equality}
\end{equation}
or by imposing a stronger radiation-dominance condition such as
\begin{equation}
\frac{\rho_r}{\rho_{\phi}+\rho_r}\geq0.99.
\label{eq:radiation_dominance_condition}
\end{equation}

The perturbative description above is not expected to encompass all possible
post-inflationary dynamics. In particular, coherent inflaton oscillations can
undergo parametric amplification and fragmentation during a nonperturbative
preheating stage, which may generate localized overdense configurations such
as oscillons. Such effects have recently been studied in the presence of an
Einstein--scalar--Gauss--Bonnet interaction~\cite{WaemingEtAl2026}. For
moderate Gauss--Bonnet corrections that remain within the regime of validity
of the effective field theory, the properties of the resulting oscillons were
found to change only mildly. For larger couplings, however, the high local
curvature generated during oscillon formation can drive the system outside
the validity regime of the truncated effective theory, so that nonlinear
backreaction can no longer be neglected.

Our calculation does not resolve the microscopic preheating dynamics or the
formation of inhomogeneous structures. It constrains the integrated expansion
history between the end of inflation and the onset of radiation domination.
If a nonperturbative stage is sufficiently short, or if its effect on the
coarse-grained expansion can be encoded in an effective $w_{\rm re}(N)$, then
its influence enters the matching equations through the e-fold-averaged
quantity $\overline{w}_{\rm re}$. Consequently, at fixed
$\overline{w}_{\rm re}$ and fixed $\rho_{\rm end}$, the effective matching
values of $N_{\rm re}$ and $T_{\rm re}$ are unchanged.

Efficient fragmentation or a long-lived oscillon population can, however,
modify the averaged equation of state, delay radiation domination, and shift
the inferred $N_{\rm re}$ and $T_{\rm re}$. If the localized curvature
becomes large enough that the Gauss--Bonnet sector leaves its weak-coupling or
effective-theory regime, the homogeneous effective description is no longer
reliable. Accordingly, the present treatment assumes that nonperturbative
resonance, fragmentation, and oscillon backreaction are either subdominant or
can be represented through the effective $w_{\rm re}(N)$, that a
coarse-grained FLRW description remains valid, and that the Gauss--Bonnet
corrections remain within their regime of validity. Testing these assumptions
quantitatively for the present $\mu(\phi,X)$ coupling requires a dedicated
nonlinear treatment and is beyond the scope of the present work.

It should therefore be emphasized that the thermal-history matching relation
constrains only the integrated quantity
\begin{equation}
\int_{0}^{N_{\rm re}}w_{\rm re}(N)\,dN,
\end{equation}
or equivalently \(\overline{w}_{\rm re}\), rather than the detailed functional
form of \(w_{\rm re}(N)\). Accordingly, the figures and tables presented in the
preceding subsections remain valid when their selected constant values are
interpreted as e-fold-averaged equation-of-state parameters. Distinguishing
between different time-dependent histories with the same average would require
additional microscopic assumptions, an explicit numerical treatment of the
coupled inflaton--radiation system, or an independent observable sensitive to
the reheating evolution, such as the primordial gravitational-wave spectrum
\cite{GhoshGhosh2025}.
Within the effective approach adopted here, the averaged equation-of-state
interpretation provides the most direct extension of the existing numerical
analysis without introducing additional model-dependent reheating parameters.

\section{Summary and conclusions}
\label{sec:conclusions}

In this work, we studied effective reheating in a
scalar--Gauss--Bonnet inflationary model with the
phase-space-dependent coupling
$\lambda_{\rm GB}\mu(\phi,X)$. The field dependence of the coupling
is localized by a compact bump, while its explicit kinetic
dependence is described by a bounded gate. The modified
inflationary background determines the pivot-scale quantities
$H_\star$ and $c_{s\star}$ and the effective energy density
$\rho_{\rm end}$, which enter the reheating duration
$N_{\rm re}$ and reheating temperature $T_{\rm re}$ through the
thermal-history matching relation.

The fixed-pivot analysis first shows how the overall Gauss--Bonnet strength $\lambda_{\rm GB}$ and the kinetic-gate amplitude $g_{_X}$ affect
reheating for the representative parameter sets. The choices
used in the two scans are listed in
Table~\ref{tab:fixed-pivot-scan-setup}, while the corresponding
values of $H_\star$, $c_{s\star}$, and $\rho_{\rm end}$ are given
in Tables~\ref{tab:lambda-background-quantities} and
\ref{tab:gx-background-quantities}. These background quantities
explain the reheating results reported in
Tables~\ref{tab:lambdaGB-reheating-scan} and
\ref{tab:gx-reheating-scan}. For the representative parameter
sets, increasing either $\lambda_{\rm GB}$ or $g_{_X}$ gives a longer
reheating stage and a lower final reheating temperature.

The additional tests in
Tables~\ref{tab:lambda-alternative-benchmarks} and
\ref{tab:gx-alternative-benchmarks} show that the magnitude of these
effects depends on the other model parameters. The effect of
$\lambda_{\rm GB}$ becomes small when the coupling is more
localized or when the E-model potential has a stronger role near
the end of inflation. The effect of $g_{_X}$ also becomes weak when
the kinetic gate is strongly bounded and the overall
Gauss--Bonnet interaction is small. The reheating behavior is
therefore set by the complete parameter choice, rather than by one
varied parameter alone.

The same parameter sets are compared with the CMB constraints in
Figs.~\ref{fig:lambda-cmb-contour},
\ref{fig:gx-cmb-contour}, and
\ref{fig:alternative-benchmark-contour}. The representative and
alternative $\lambda_{\rm GB}$ benchmarks remain in the low-$r$
region and close to the lower-$n_s$ part of the displayed
contours. The kinetic benchmarks lie at larger $n_s$, with the
alternative kinetic benchmark showing the closest agreement with
the Planck and ACT regions considered here.

The reheating constraints of these four benchmarks are shown in
Figs.~\ref{fig:lambda-reheating-curves},
\ref{fig:gx-reheating-curves},
\ref{fig:alternative-lambda-reheating}, and
\ref{fig:alternative-kinetic-reheating}. The corresponding ranges
of $N_{\rm re}$ and $T_{\rm re}$ that satisfy the adopted CMB range,
$N_{\rm re}\geq0$, and the BBN temperature condition are listed in
Tables~\ref{tab:lambda-reheating-ranges} and
\ref{tab:kinetic-reheating-ranges}. All four representative values
of the effective reheating equation-of-state parameter considered
in this work retain allowed ranges. In the generalized
varying-equation-of-state description, these values can equivalently
be interpreted as the e-fold-averaged parameters
$\overline{w}_{\rm re}$.

The representative and alternative $\lambda_{\rm GB}$ benchmarks
give similar reheating ranges. Their main differences appear for
$\overline{w}_{\rm re}=-1/3$ and
$\overline{w}_{\rm re}=0$, while the stiff branches remain close in
the two cases. The kinetic benchmarks show a clearer difference,
especially for $\overline{w}_{\rm re}=-1/3$ and
$\overline{w}_{\rm re}=0$. For the stiff averaged equations of
state, the BBN temperature bound places the stronger restriction,
and the $\overline{w}_{\rm re}=2/3$ branches have the narrowest
allowed ranges.

Overall, Tables~\ref{tab:fixed-pivot-scan-setup}--%
\ref{tab:kinetic-reheating-ranges} and
Figs.~\ref{fig:lambda-cmb-contour}--%
\ref{fig:alternative-kinetic-reheating} give a consistent picture.
The inflationary observables and the reheating constraints both
depend on the potential, the field-space localization of the
coupling, the kinetic gate, and the overall Gauss--Bonnet strength.
The CMB comparison tests the values of $n_s$ and $r$, while the
reheating analysis gives the corresponding allowed ranges of
$N_{\rm re}$ and $T_{\rm re}$ for representative values of the
effective, or equivalently e-fold-averaged, reheating
equation-of-state parameter $\overline{w}_{\rm re}$.

The analysis was also generalized to allow the reheating
equation-of-state parameter to vary during the post-inflationary
evolution. In this case, the thermal-history matching relation
depends on the e-fold-averaged quantity
\begin{equation}
\overline{w}_{\rm re}
=
\frac{1}{N_{\rm re}}
\int_{0}^{N_{\rm re}}w_{\rm re}(N)\,dN.
\label{eq:average-wre-conclusion}
\end{equation}
Consequently, the numerical framework and the results presented in
the preceding figures and tables remain applicable without
recalculating the inflationary background. The constant values of
$w_{\rm re}$ used in the numerical analysis can equivalently be
interpreted as representative values of
$\overline{w}_{\rm re}$. Therefore, any time-dependent reheating
history with the same e-fold-averaged equation of state produces
the same values of $N_{\rm re}$ and $T_{\rm re}$ within the
effective thermal-history matching framework adopted here.

It should nevertheless be emphasized that the thermal-history matching
relation constrains only the e-fold-averaged parameter
$\overline{w}_{\rm re}$ and cannot determine the detailed evolution of
$w_{\rm re}(N)$. Distinguishing between different reheating histories
with the same e-fold average would require a more microscopic treatment
involving explicit inflaton decay channels, particle production,
preheating dynamics, and the coupled evolution of the inflaton and
radiation sectors. Such an analysis, possibly supplemented by
observables sensitive to the detailed reheating history, such as the
primordial gravitational-wave spectrum
\cite{GhoshGhosh2025}, provides a natural direction for future work.

Nonperturbative preheating provides an additional limitation of the effective
description adopted here. In particular, inflaton fragmentation and oscillon
formation may modify the coarse-grained equation of state and the duration of
the transition to radiation domination. Our reheating constraints therefore
remain directly applicable to such histories only when their effects can be
encoded in the same e-fold-averaged $\overline{w}_{\rm re}$ and when the
inhomogeneous Gauss--Bonnet dynamics remains within the validity regime of the
effective theory. A dedicated nonlinear analysis of preheating and oscillon
formation for the present $\mu(\phi,X)$ coupling would be required to
determine this regime quantitatively.

\begin{acknowledgments}
Use of AI-assisted technologies. In preparing this manuscript, the authors used AI-assisted technologies to improve language quality, grammar, and \LaTeX{} typesetting. Every suggestion generated by these tools was critically evaluated and, where necessary, modified by the authors. The authors retain full responsibility for the originality, accuracy, and integrity of the intellectual content, including all analyses, interpretations, and conclusions.

\end{acknowledgments}

\end{document}